\documentclass[prl,reprint,twocolumn,
aps,amssymb,graphicx,floatfix,amsmath,superscriptaddress,showpacs,longbibliography]{revtex4-1}
\usepackage{bm}
\usepackage{float}
\usepackage{graphicx}
\usepackage{color}
\usepackage{amsfonts}
\usepackage{amsmath}
\usepackage{textcomp}
\usepackage{subcaption}
\usepackage{hyperref}
\usepackage{comment}
\usepackage{dsfont}
\hypersetup{colorlinks=true,linkcolor=blue,anchorcolor=blue,citecolor=blue,filecolor=blue,urlcolor=blue,bookmarksnumbered=true,pdfview=FitB}
\newcommand{\e}{\epsilon}

\newcommand{\beq}{\begin{equation}}
\newcommand{\eeq}{\end{equation}}
\newcommand{\bea}{\begin{eqnarray}}
\newcommand{\eea}{\end{eqnarray}}

\newcommand{\nn}{\nonumber}

\newcommand{\re}{\text{Re}}
\newcommand{\im}{\text{Im}}

\newcommand{\bk}{{\textbf k}}

\newcommand{\bq}{{\textbf q}}
\newcommand{\br}{{\textbf r}}

\newcommand{\bwt}{\begin{widetext}}
\newcommand{\ewt}{\end{widetext}}
\newcommand{\bse}{\begin{subequations}}
\newcommand{\ese}{\begin{subequations}}

\newcommand{\cm}{\,\text{cm}^{-3}}

\newcommand{\ve}{\varepsilon}
\newcommand{\te}{\tilde\epsilon}

\begin{document}
\title
{
Quasiparticle and Nonquasiparticle Transport in Doped Quantum Paraelectrics}
\author{Abhishek Kumar}
\affiliation{University of Florida, Gainesville, Florida, 32611, USA}
 \author{Vladimir I. Yudson}
 \affiliation{Laboratory for Condensed Matter Physics, National Research University ``Higher School of Economics'', 
20 Myasnitskaya St., Moscow, 101000 Russia}
\affiliation{Russian Quantum Center, Skolkovo, Moscow 143025, Russia}
\author{Dmitrii L. Maslov}
\affiliation{University of Florida, Gainesville, Florida, 32611, USA}
\date{\today}

\begin{abstract}
Charge transport in doped quantum paralectrics (QPs) presents a number of puzzles, including a pronounced $T^2$ regime in the resistivity. We analyze charge transport in a QP within a model of electrons 
coupled to a soft transverse optical (TO) mode via a two-phonon mechanism. For $T$ above the soft-mode frequency but below some characteristic scale ($E_0$), the resistivity scales with the
occupation
 number of phonons squared, i.e., as $T^2$. The $T^2$ scattering rate does not depend on the carrier number density and is not affected by a crossover between degenerate and non-degenerate regimes, in agreement with the experiment.   Temperatures higher than $E_0$ correspond to a non-quasiparticle regime, which we analyze by mapping the Dyson equation onto a problem of supersymmetric  quantum mechanics. The combination of scattering by two TO phonons and by a longitudinal optical mode
explains the data quite well.  \end{abstract}

\maketitle

Quantum paraelectrics (QPs)  are materials close to a ferroelectric transition but never quite making it because of zero-point motion which disrupts symmetry breaking~\cite{mueller,Chandra:2017,Stemmer2018,Collignon:2019}. This group includes several perovskites, e.g., SrTiO$_3$ (STO), 
  KTaO$_3$ (KTO), 
 and  EuTiO$_3$ (ETO),
 and a number of rock salts, e.g., PbTe. Electron transport in doped QPs is very much different from that in doped semiconductors. To begin with, a very large static dielectric constant ($\sim 25,000$ in STO)
translates into a micron-long Bohr radius. Consequently, 
conduction in QPs sets in at very low doping, e.g., at few times $10^{15}\cm$ carriers in STO \cite{spinelli}, and is prominently metallic above $10^{17}\cm$.  In the metallic regime, the resistivity increases by several orders of magnitude from the helium to room temperatures, exceeding formally the Mott-Ioffe-Regel (MIR) limit around 100 K \cite{kamrannat}.
A very intriguing observation is a prominent $T^2$ scaling of the resistivity observed in  STO \cite{kamranscience,mikheev,leighton_sp},
KTO \cite{Sakai}, and ETO \cite{Engelmayer2019}. Normally, a $T^2$ resistivity is associated with the Fermi-liquid (FL) behavior.
However, a $T^2$ resistivity  in QPs is observed already at very low doping, when umklapp scattering is forbidden  and only the lowest conduction band is occupied \cite{kamranscience}, and straddles a number of relevant energy scales, such as the plasma frequency and the Fermi energy ($E_F$). In addition, the $T^2$ scattering rate depends only weakly on the electron number density, $n$ \cite{kamranscience,mikheev}. All of the above contradicts the interpretation of the $T^2$ behavior in terms of the FL theory \cite{mikheev,maslov2017,Swift2017,Stemmer2018}.

In this Letter, we discuss the model of electrons interacting with a soft transverse optical (TO) mode, which is a defining feature of QPs. As temperature is lowered, the frequency of the TO mode decreases, indicating the tendency to a ferroelectric transition, but eventually saturates at a small but finite value (as low as $\omega_0\approx 1$\,meV for the $E_u$ mode in STO \cite{Vogt1995,Hehlen1999,yamanaka:2000}). For electrons near the Brillouin zone center, single-TO phonon scattering is suppressed in a single-band system and in the absence of spin-orbit interaction \cite{Balatsky2018,Ruhman_comment,Ruhman:2016,volkov:2020,GASTIASORO2020,gastiasoro_b:2020}, and the lowest-order process involves two TO phonons (2TO)\cite{ngai:1974,levanyuklong,levanyukshort}.

We show that the model is characterized by a material-dependent energy scale, $E_0$, separating the regimes of quasiparticle and nonquasiparticle transport, at lower and higher $T$, respectively.  
(In STO, $E_0$ is on the order of $100$\,K).
For $\omega_0\ll T\ll E_0$, the TO mode is in the classical regime, and  a $T^2$ term in the resistivity arises simply from the square of the phonon occupation number. A unique feature of the 2TO mechanism is that the quasiparticle scattering rate, $1/\tau\sim T^2/E_0$, does not depend on the electron energy. This explains why the observed $T^2$ scattering rate  depends on $n$ only weakly  for $T\ll E_F$ and does not  exhibit a crossover at $T\sim E_F$.

For $T\gg E_0$, the quasiparticles are not well-defined. By mapping the Dyson equation for the self-energy onto an exactly soluble problem of supersymmetric
quantum mechanics, we show that transport in this regime is dominated by severely {\em off-shell} electrons.
In this regime, the resistivity scales as $T^{3/2}$
and violates the MIR limit.

Finally, we show that a more realistic model, which incorporates the $T$ dependence of the TO soft mode and also includes scattering by longitudinal optical (LO) phonons, explains the experimental data, {\em if} the freeze-out of TO phonons for $T<\omega_0$ is ignored. We discuss the advantages and shortcomings of the 2TO model and propose a number of experiments that can falsify it.

\begin{figure}
\centering
\includegraphics[scale=0.4]{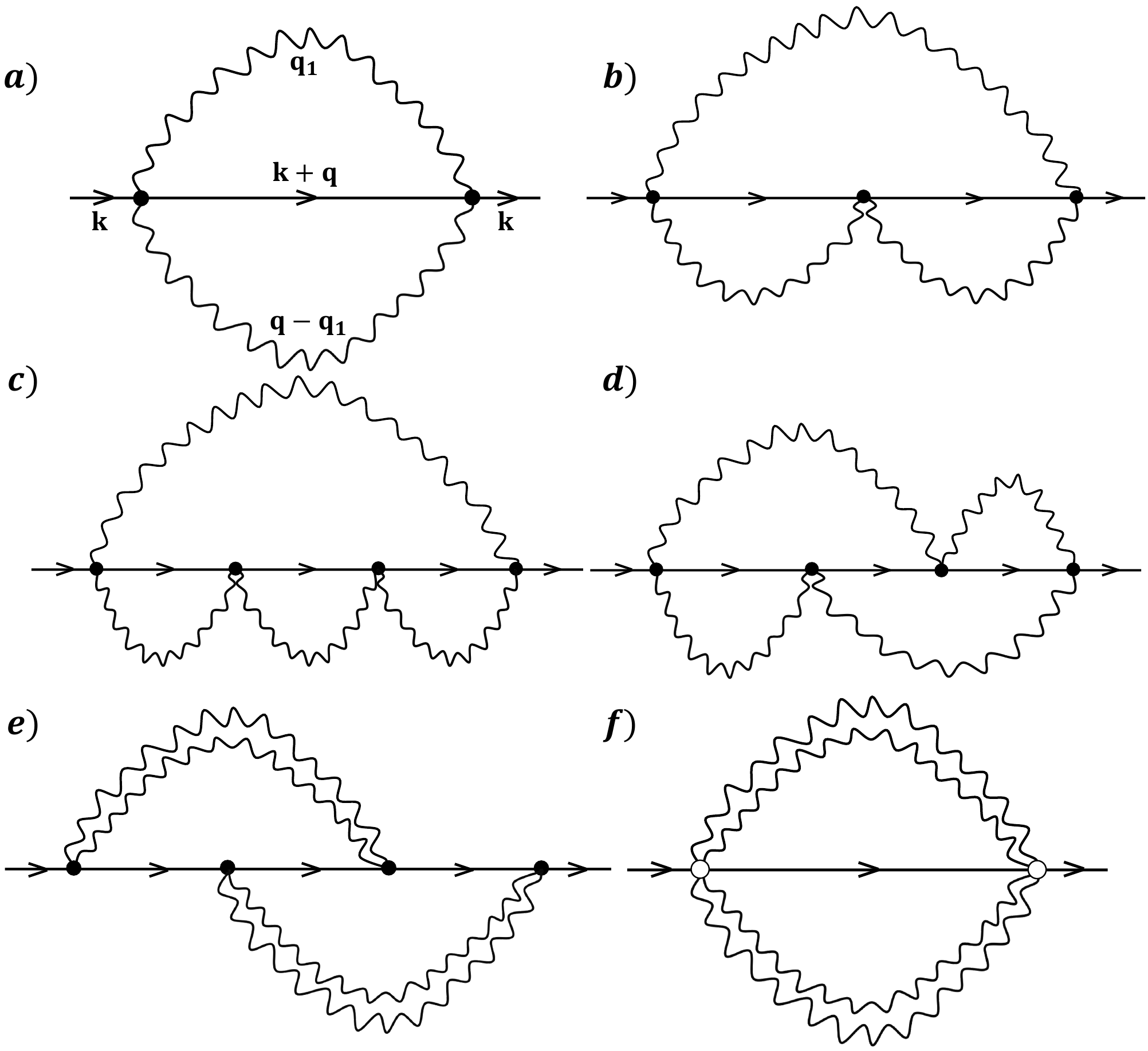}
\caption{\label{fig:diag} Diagrams for the electron self-energy due to scattering by TO phonons. {\em a}) Two-loop two-phonon diagram; {\em b}) and {\em c}) three- and four-loop ``umbrella'' diagrams without crossings; {\em d}) and {\em e}) examples of diagrams with crossings; {\em f}) four-phonon diagram resulting from adding a ${\bf P}^4$ term to Eq.~(\ref{ham}).}
\end{figure}
We consider 3D electrons coupled to an $O(3)$ electric polarization ${\bf P}(\br)$, produced by TO phonons.  Because $\boldsymbol{\nabla}\cdot{\bf P}=0$,
single-phonon coupling is forbidden and the Hamiltonian starts with a two-phonon term \cite{levanyuklong,levanyukshort}:
\beq
\label{ham}
H_{2\text{TO}} =
 \frac{g_2}{2} \int d^3r 
{\bf P}
^2 (\br)
\psi^\dagger(\br) \psi(\br),
\eeq
where $g_2$ is the coupling constant (with units of the volume)
\footnote{Our definition of $g_2$ differs by a factor of 2 from that in Refs.~\cite{levanyuklong,levanyukshort}. Our choice eliminates a combinatorial factor of $2^{N-1}$ in the ``umbrella'' diagrams  {\em a}-{\em c} in Fig.~\ref{fig:diag} with $N^{\text{th}}$ vertices, which is equal to the number of ways a correlation function $<~\Pi_{n=1}^N P^2({\bf r}_n)>$ can be partitioned into binary averages.}.
Other than allowing for TO modes, we treat the material as isotropic. For a TO mode with dispersion $\omega^2_\bq=\omega_0^2+s^2q^2$ and polarization 
${\bf e}^{a}_\bq$,
\bea
\label{pol}
{\bf P}
(\br)
=\sum_{\bq,a} \frac{{\bf e}^a_\bq}{\sqrt{V}}
A_\bq b_\bq e^{i \bq\cdot\br} +\text{h.c.}
\eea
where $A^2_\bq=\left[\ve_0(\bq) - \ve_\infty\right]\omega_\bq/4\pi$ \cite{kittelsolids},  
the sum over $a=1,2$ accounts for two (degenerate) branches of the TO mode, 
$\ve_0(\bq)$ and $\ve_\infty$ are the static and high-frequency limits of the dielectric function, respectively, and $b_\bq $ is the bosonic annihilation operator.   The diagrams for the electron self-energy are shown in Fig.~\ref{fig:diag}, where the solid and wavy lines denote the (Matsubara) electron and phonon Green's functions, $G(\bk,\e_m)$ and $D(\bq,\omega_m)$, respectively,
and solid dots denote the electron-2TO-phonon vertex 
$ \Gamma_{\alpha\beta}(\bq)= g_2A_\bq^2(\delta_{\alpha\beta}-q_\alpha q_\beta/q^2)$.
Phonons will be treated as bare ones, hence $D(\bq,\omega_m)=-2\omega_\bq/(\omega_m^2+\omega_\bq^2)$.

We now focus on the classical regime, 
when phonons can be treated as static ``thermal disorder''\footnote{The classical regime sets in for $T\gg \max\{T_{\text{BG}},\omega_0\}$, where $T_{\text{BG}}=2k_Fs$ is the Bloch-Gr{\"u}neisen temperature. A solution of the Boltzmann equation for 2TO scattering \cite{SM} shows that the actual crossover temperature between the classical and Bloch-Gr{\"u}neisen regimes is numerically small ($0.25 T_{\text{BG}}$) and below $\omega_0$ at $n$ considered in this paper.}, which corresponds to setting $\omega_m=0$ in the phonon lines. After analytic continuation $i\epsilon_m\to \epsilon+i 0^+$, Fig.~\ref{fig:diag}({\em a}) yields
\bea
\label{sigma}
\Sigma(\bk,\epsilon)= \int_{\bq}
G(\bk+\bq,\epsilon) U(\bq),
\eea
 where the correlation function of thermal disorder is 
 \bea
 U(\bq)=2{T^2}\hspace{-0.05in}\int
\frac{d^3q_1}{(2\pi)^3} \sum_{\alpha\beta} 
 \frac{\Gamma_{\alpha\beta}(\bq_1)}{\omega_{\bq_1}}\frac{\Gamma_{\beta\alpha}(\bq-\bq_1)}{\omega_{\bq-\bq_1}}. \eea
Other diagrams can be treated in a similar manner.

We also assume for now that the material is very close to the quantum-critical point, so that the gap in the phonon dispersion
 can be neglected, i.e., $\omega_\bq=sq$.
   Neglecting also $\epsilon_\infty$ compared to $\ve_0(\bq)$ and excluding  $\ve_0(\bq)$ via the Lyddan-Sachs-Teller (LST) relation, $\ve_0(\bq)=\Omega_0^2/\omega_\bq^2$, and integrating over $\bq_1$, we obtain
   \bea
   U(\bq)=\frac{3\pi}{2m^*q} \frac{T^2}{E_0}\;\text{with}\;E_0\equiv \frac{64\pi^3 s^4}{m^*g_2^2 \Omega_0^4},\label{E0}
   \eea
where $E_0$ is characteristic energy scale of the model.
The $1/q$ scaling of  $U(\bq)$ (or $1/r^2$ scaling is real space) will be crucial in what follows.

For $T\ll E_0$, thermal disorder is weak. This is the quasiparticle regime, when diagram  Fig.~\ref{fig:diag}({\em a}) with $G$ replaced by its free-electron form, $G_0(\bk,\epsilon)=(\epsilon-\xi_\bk+\mu+i0)^{-1}$ with $\xi_\bk=k^2/2m^*$, gives the leading-order result. Accounting also for a transport correction, we obtain the standard result for the transport scattering rate
\bea
\frac{1}{\tau} = 
2\pi\int \frac{d^3 q}{(2\pi)^3} \delta(\xi_{\bk+\bq}-\xi_\bk)U(\bq)(1-\cos\theta),\label{tau}
\eea
where $\theta$ is the angle between $\bk$ and $\bk+\bq$. (The difference between the quantum and transport rates is insignificant because our thermal disorder is relatively short-ranged; as a result,  the two rates differ only by a factor of $2/3$.)

In general, $\tau$ depends on the electron energy, $\xi_\bk$, via the electron density of states. This is the reason why, for example, the resistivity of a semiconductor due to acoustic phonon scattering scales as $T$ for $T\ll E_F$ and as $T^{3/2}$ for $T\gg E_F$. Our case of $U(\bq)\propto 1/q$ is, however,  special: the $1/q$ factor cancels out with the density of states, and the result does not depend on $\xi_\bk$. 
Evaluating also Fig.~\ref{fig:diag}({\em b}) and ({\em c}), we obtain 
\bea
\frac{1}{\tau}
=\frac{T^2}{E_0}-1.24\frac{T^3\sqrt{m^*}}{k E^{3/2}_0}+\mathcal{O}\left( \frac{T^5m^{*3/2}}{k^3 E_0^{5/2}}\right).
\label{subP}
\eea

The leading term in Eq.~(\ref{subP}) is the most relevant one for the experiment: because it does not depend on $\xi_\bk$, its thermal average does not depend on the statistics of charge carriers, and the corresponding resistivity
\bea
\rho=\frac{m^*}{ne^2}\frac{T^2}{E_0}\label{rhoT2}
\eea scales as $T^2$ regardless of whether $T$ is lower or higher than $E_F$. 
From the data \cite{kamran_thanks}, we extract $E_0=209$\,K in STO at $n=4\times 10^{17}\cm$. Using the known parameters of the phonon spectrum \cite{Yamada} ($s=6.6\times 10^{5}$\,cm/s and $\Omega_0=194.4$\,meV) and  $m^*=1.8m_0$ \cite{kamraneffectivemass},  we find that $E_0=209$\,K corresponds to
$g_2= 
0.60
 a^3_0$, where  $a_0=3.9$\,\AA\ is the STO lattice constant. 
This is close to an earlier estimate  \cite{levanyukshort,levanyuklong} of $g_2 
=1.0 a^3_0$.

Strong thermal disorder ($T\gg E_0$) corresponds to a non-quasiparticle regime.
Since $E_F\ll E_0$ for the relevant range of electron number densities,  we will consider the non-degenerate case only. According to Eq.~(\ref{subP}), $1/\tau$ becomes comparable to the electron energy $(T)$ at $T\sim E_0$. If TO scattering is treated as purely elastic, the condition  $T\tau\sim 1$ should indicate the onset of Anderson localization. However, small but finite energy transfers give rise to dephasing, which turns out to be strong enough to prevent localization. 
Indeed, in a typical scattering event electron energy is changed by $\delta\epsilon\sim k_Ts$, where $k_T\sim \sqrt{m^*T}$ is the thermal electron momentum. This corresponds to diffusion along the energy axis with a diffusion coefficient ${\mathcal D}_\epsilon\sim (\delta\epsilon)^2/\tau\sim m^*s^2T^3/E_0$. The phase-breaking time $\tau_{\phi}$ can be estimated from  the condition 
  that the phase accumulated during $\tau_{\varphi}$ is on the order unity, i.e.,
$\Delta\phi=\Delta\epsilon\tau_\phi=({\mathcal D}_\epsilon\tau_{\phi})^{1/2}\tau_{\phi}\sim 1$ \cite{Altshuler:1982} or $\tau_{\phi}\sim (E_0/m^*s^2)^{1/3}/T$. We see that $\tau_{\phi}$ becomes comparable to the elastic time $\tau\sim E_0/T^2$ at $T\sim T_{\phi}=(m^*s^2/E_0)^{1/3}E_0\ll E_0$, i.e., already in the quasiparticle regime, 
and it is reasonable to assume that localization can be neglected for all $T>T_\phi$.

We now find the self-energy self-consistently from 
Dyson equation (\ref{sigma}). Relabeling $\bq=\bk-\bk'$ and integrating over the angle between $\bk$ and $\bk'$, we obtain 
\bea
\Sigma(\xi,\epsilon)=\lambda \int_0^\infty d\xi' K(\xi,\xi') \frac{1}{\tilde\e-\xi
-\Sigma(\xi',\e)},\label{integ}
\eea
where $\tilde\e=\e+\mu$,  $\lambda=3T^2/4\pi E_0$, $\xi\equiv\xi_{\bk}$, $\xi'\equiv \xi_{\bk'}$, and $K(\xi,\xi')=\sqrt{\xi'/\xi} \Theta(\xi-\xi')+\Theta(\xi'-\xi)$.
At weak coupling ($T\ll E_0)$,  when the Green's function can be replaced by its free-electron form, $\im\Sigma(\xi,\e)=-\pi\lambda\Theta(\tilde\e)\left[ \Theta(\xi-\tilde\e)\sqrt{\tilde\e/\xi}+\Theta(\tilde\e-\xi)\right]$ is non-zero only above the bottom of the band \footnote{The mass-shell limit of the last formula reproduces the first term in Eq.~(\ref{subP}), up to a transport correction of $2/3$.}. We will now show that at strong coupling ($T\gg E_0$)  the threshold in $\im\Sigma(\xi,\e)$ 
moves from $\tilde\e=0$ to a finite value which depends on the coupling constant.
This is an essentially non-perturbative effect that defines transport in the non-quasiparticle regime.

If a threshold does exist, $\im\Sigma(\xi,\epsilon)$ must be small right above the threshold. Therefore, Eq.~(\ref{integ}) can be expanded in 
$\gamma_\epsilon(\xi)\equiv -\im\Sigma(\xi,\epsilon)$.
On the other hand, $\re\Sigma(\xi,\epsilon)$ is expected to be regular 
near the threshold 
and to depend on $\xi$ only weakly, so it can be absorbed into the chemical potential. (Using Kramers-Kronig relation, one can show that Re$\Sigma$ depends on $\xi$ and $\epsilon$ only logarithmically \cite{SM})
Assuming that relevant $\tilde\e<0$, we expand the imaginary part of Eq.~(\ref{integ}) in $\gamma_{\epsilon}(\xi)$ as:
\bea
\gamma_{\epsilon}(\xi)=\lambda \int_0^\infty d\xi' K(\xi'/\xi)\left[\frac{\gamma_{\epsilon}(\xi')}{(\tilde\e-\xi')^2}-\frac{\gamma_{\epsilon}^3(\xi')}{(\tilde\e-\xi')^4}\right].\label{cubic}
\eea
At first, we drop the cubic term. The linearized integral equation can be transformed into a ``zero-energy Schroedinger equation'' for $\varphi_{\epsilon}(\xi)\equiv\xi^{3/4}\gamma_{\epsilon}(\xi)$   \cite{SM}: \bea
\left[- \partial_\xi^2+V(\xi)\right]\varphi_{\epsilon}(\xi)=0;\;V(\xi)=-\left[\frac{3}{16\xi^2}+\frac{\lambda}{2\xi(\tilde\e-\xi)^2}\right].\nn\\
 \label{Sch}
 \eea  
The threshold is defined as the smallest value of $\tilde\e$ at which the zero-energy Schroedinger equation has a non-trivial solution, which is guaranteed to be the case if 
the Hamiltonian, $H_{\varphi}=-\partial_\xi^2+V(\xi)$, is supersymmetric (SUSY) \cite{Cooper:1995}.  This means  that $H_{\varphi}$ can be written as $H_{\varphi}=Q^\dagger Q$, where $Q=\partial_\xi+W(\xi)$,
$Q^\dagger=-\partial_\xi+W(\xi)$, and $W(\xi)$ is a superpotential satisfying the Riccati equation $W^2(\xi)-W'(\xi)=V(\xi)$. It can be verified \cite{SM} that the Riccati equation is solved by
$W(\xi)=-3/4\xi+1/2(\xi-\tilde\e)$ if $\tilde\e/\lambda=-2/3$, which is the condition for $H_{\varphi}$ to be of the SUSY type.  This implies that the threshold in the self-energy is located at $\tilde\e=-2\lambda/3\equiv -\e_0$, while the first-order equation $Q\varphi_{\epsilon}=0$ yields $\gamma_{\e}(\xi)=\xi^{-3/4}\varphi_{\epsilon}(\xi)=C(\tilde\e)/\sqrt{\xi+\e_0}$. The function  $C(\tilde\e)$ is found by substituting the last equation in Eq.~(\ref{cubic}) and retaining the cubic term.
The final result for $\im\Sigma$ near the threshold reads
 \bea
 \im\Sigma(\xi,\e)=-\sqrt{(\e_0+\tilde\e)\e_0}\; S(\xi/\e_0),\label{gamma}
 \eea
 where
\bea
S(x)=\left[\frac{42(x+1)(2x+3)}{16x^3+56x^2+70x+35}\right]^{1/2}.\label{g}
\eea

Note that what is relevant for the observables is the threshold in $\tilde\e$ rather than in $\e$ itself. Nevertheless, we need to determine $\mu$, as it is not guaranteed that at strong coupling electrons are still in the non-degenerate regime. 
Imposing the constraint of fixed number density, we find  $\mu=-\e_0-(3T/2)\ln(T/E_F)$ \cite{SM}.
Because $\mu<0$ and $|\mu|\gg T$,
we are indeed in the non-degenerate regime.

 To find the resistivity in the nonquasiparticle regime, we ignore the vertex corrections of both ladder and Cooperon types for reasons given above. Then
\bea
\rho=
\frac
{3 m^* T}{2e^2 n}
\frac{
\int^\infty_{-\infty} d\te e^{-\te/T}\int^\infty_0 d\xi N(\xi)(-)\im G(\xi,\e)}
{\int^\infty_{-\infty} d\te e^{-\te/T}\int^\infty_0 d\xi N(\xi) \xi  \left[\im G(\xi,\e)\right]^2},
\nn\\\label{kubo}
 \eea
where $N(\xi)=m^{*3/2}\sqrt{2\xi}/\pi^2$ is the density of states.  The 
numerator in Eq.~(\ref{kubo}) comes from  the relation between the chemical potential and number density.
The lower limit in the $\te$-integrals is $-\e_0$, and the Boltzmann factor $e^{-\tilde\e/T}$
is exponentially large near $-\e_0$. Therefore, the $\te$ integrals come from the near-threshold region,
where the self-energy is given by Eqs.~(\ref{gamma}) and (\ref{g}). Substituting these forms into Eq.~(\ref{kubo}), we obtain
\bea
\rho=5.6\frac{m^*}{ne^2}\sqrt{T\e_0}\propto T^{3/2}.
\label{rho_high}
\eea

\begin{figure}[H]
\includegraphics[scale=0.355]{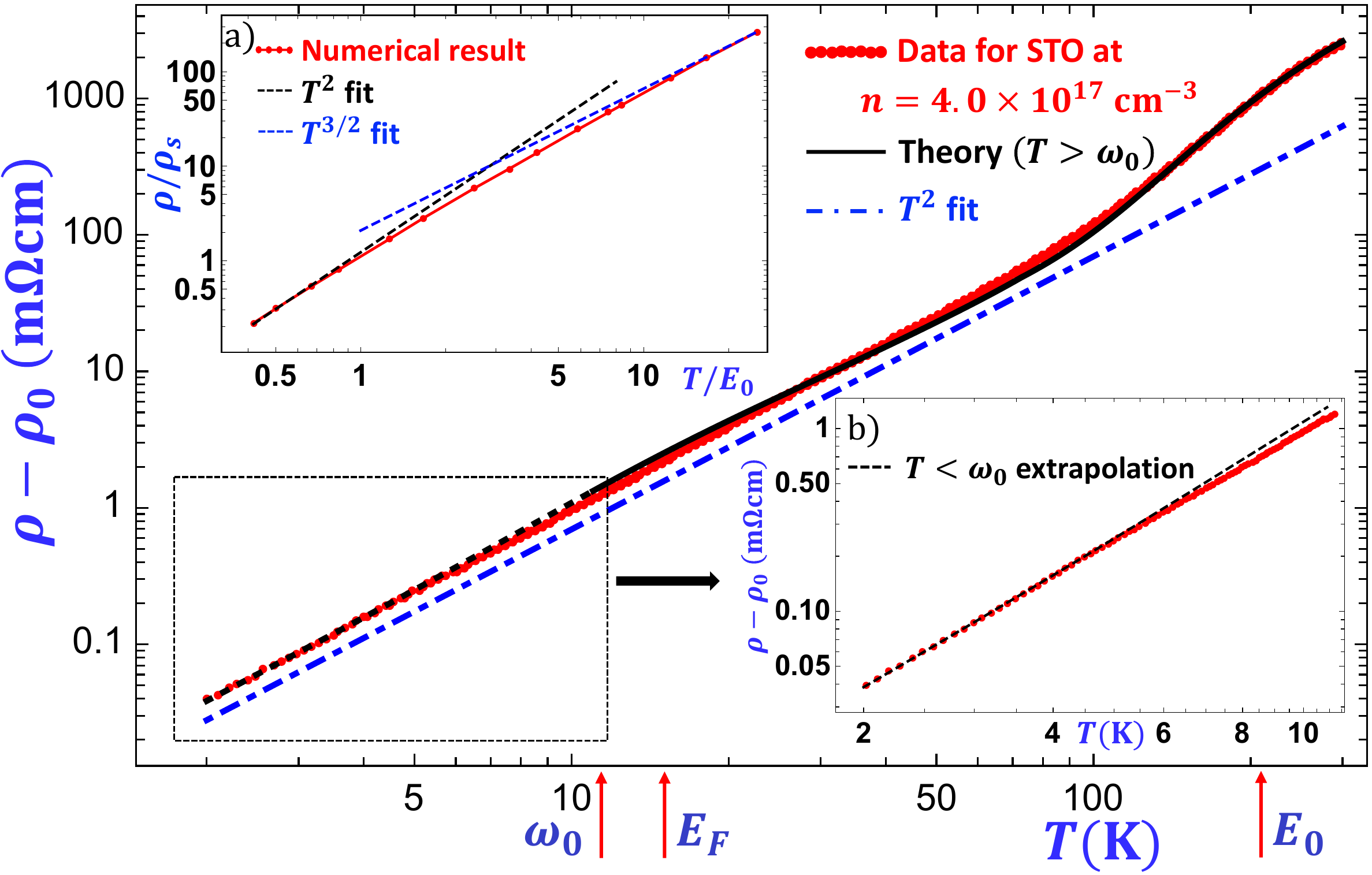}
\caption{\label{fig:LOTOmod} Main panel: Resistivity (minus the residual value $\rho_0$) of SrTiO$_3$ \cite{kamran_thanks} (points, red) vs theory (solid, black), which includes scattering by two TO phonons and by the 58 meV LO phonon. An extrapolation of the theory to the regime of $T < \omega_0$ regime is shown by the dashed line. The dash-dotted line is a $T^2$ fit to the data (shifted for clarity). Insets: (a) The temperature dependence of the resistivity predicted by the 2TO model, obtained by a numerical solution of Eqs.~(\ref{integ}) and (\ref{kubo}), along with the fits to the asymptotic results. Here, $\rho_s = m^*E_0/ne^2$. Inset b) An enlargement on the low-temperature region of the main panel.} \end{figure}

Despite the Drude-like appearance of Eq.~(\ref{rho_high}), its physical content is very different because transport in this regime is controlled by  off-shell electrons with $\tilde\e\approx -\e_0$ and  $\xi\sim \e_0$. However, if one still chooses to interpret Eq.~(\ref{rho_high}) in a Drude-like way,  the corresponding scattering time $\tau_D\sim E_0^{1/2}/T^{3/2}$ 
is shorter than the Planckian bound, $\tau_P=1/T$, for $T\gg E_0$. 
In Supplemental Material \cite{SM}, we show that the analytic results in Eqs.~(\ref{rhoT2}), (\ref{gamma}) and (\ref{rho_high}) are confirmed by a numerical solution of Eq.~(\ref{integ}). In particular,
the inset in Fig.~\ref{fig:LOTOmod} shows the resistivity obtained by substituting a numerical solution of  Eq.~(\ref{integ}) into Eq.~(\ref{kubo}). 

We now discuss briefly the role of other diagrams in Fig.~\ref{fig:diag}.
For $E_F\ll T\ll E_0$, the higher-order umbrella' diagrams [Figs.~\ref{fig:diag}({\em b}), \ref{fig:diag}({\em c}), etc.], provide corrections of order $\sqrt{T/E_0}$, as specified in Eq.~(\ref{subP}). For $T\gg E_0$, it is the self-energy near the threshold that matters to transport. Near the threshold, umbrella diagrams modify scaling function $S$ in Eq.~(\ref{gamma}) but not the square-root singularity in $\im\Sigma$ as a function of $\e$ \cite{SM}. Therefore, these diagrams affect only the numerical coefficient in Eq.~(\ref{rho_high}) but not  the $T^{3/2}$ scaling of $\rho$.  Next, Fig.~\ref{fig:diag}({\em e})  is a vertex correction to Fig.~\ref{fig:diag}({\em a}), which is small by an effective Migdal parameter, $m^*s^2/E_0
\sim 0.03
$  \cite{SM}. 
Finally, diagram {\em f} describes a four-phonon process, which gives  a subleading correction to the resistivity for $T$ below the melting temperature.

We now compare the theoretical results to the data for STO, restoring the gap ($\omega_0$)  in the phonon dispersion. The $T$ dependence of $\omega_0$ is obtained by substituting the measured $\ve_0(T)$  \cite{kamran_thanks} into the LST relation [above Eq.~(\ref{E0})]. 
However, due to a partial cancelation between the $T$-dependences of $\omega_0$ and of the rms electron momentum, the $T$ dependence of $\omega_0$ does not change the results significantly  \cite{levanyuklong,levanyukshort,SM}.
The 2TO contribution to the resistivity is described by an interpolation formula which reproduces the analytic results at low and high $T$ [Eqs.~(\ref{rhoT2}) and (\ref{rho_high}), respectively],
with 2TO coupling constant $g_2$ as a fitting parameter. In the experiment, $\rho$ varies faster than $T^2$ at higher $T$: a power-law fit gives $\rho\propto T^{2.7-3}$ \cite{Wemple1965,Wemple:1966,wemple1969,kamrannat}. 
An exponent larger than 2 was conjectured to result from multi-TO-phonon scattering \cite{wemple1969}. However, we have shown 
that  TO scattering gives a slower than $T^2$ variation of $\rho$  for $T\gg E_0$ [cf.~Eq.~(\ref{rho_high})]. An alternative explanation of the faster than $T^2$ dependence is scattering by LO phonons, \cite{frederiksemob,baratoff,mikheevth,verma}. We adopt the latter model here and include scattering by the 58 meV LO mode within the Low-Pines approach \cite{lowpines}, treating the Fr\"ohlich coupling constant $\alpha$ as a fitting parameter; details of the fitting procedure are delegated to SM \cite{SM}.

On the low-$T$ side, the 2TO model should give $\rho\propto \exp(-\omega_0/T)$ for $T\ll \omega_{0}$, whereas the observed resistivity continues to scale down as $T^2$ up to the lowest $T$ measured (2 K) \footnote{While the low-$T$ resistivity exponent in oxygen-reduced STO bulk samples stays close to 2 for $n$ up to $4\times 10^{19}\cm$ \cite{Collignon:2019}, $d\rho/dT$ of La-doped thin films varies faster than $T$  for $T\lesssim 10$\,K and $n\gtrsim 5\times 10^{18}\cm$ \cite{Stemmer2018}. Similar anomalies are also observed in some Nb- and La-doped single crystals at high doping \cite{Lin_private}. It is not clear whether these anomalies indicate the onset of the Bloch-Gr{\"u}neisen/exponential behavior or are due to inhomogeneous doping.}.
Nevertheless, if we  extrapolate our 
model to the region of $T\leq \omega_0$ (where it should not be applicable), it still provides a surprisingly good fit of the data.
A fit obtained in this way is shown in Fig.~\ref{fig:LOTOmod} for $
g_2=0.92
a_0^3$ and $\alpha=2.38$ \footnote{The corresponding value of the BCS coupling constant $\lambda_{\text{BCS}}=g_2^2\Omega_0^4 m^* k_F/8\pi^4 s^3\approx 10^{-2}$ for  $n=10^{18}\cm$ and  other  parameters being the same as quoted in the text.}.
This value of $g_2$, obtained from fitting over the entire range of $T$, is slightly larger than $0.60 a_0^3$, obtained by fitting only the $T^2$ part of the data. To the best of our knowledge, no {\em ab initio}  estimate of $g_2$ is currently available and would be highly desirable. The value of $\alpha$ is higher than $\alpha\approx 0.7$ \footnote{In Refs.~\cite{eagles} and \cite{barker_1966}, the value 0.5 is actually for $\alpha(m_e/m^*)^{1/2}$. Taking effective mass into account, i.e, $m^*/m_e = 1.8$, one would get the $\alpha=0.7$ as mentioned in Ref.~\cite{frederiksemob}} extracted from infrared reflectivity \cite{eagles,barker_1966} and transport at high $T$ ($200<T<1000$\,K) \cite{frederiksemob}, but  is consistent with other transport measurements in the intermediate temperature range ($100 < T < 200$\, K)\cite{mikheevth,Mikheev:PhD}.

While we do not have a good answer to the question why the theory, extrapolated to $T<\omega_0$, still appears to describe the experiment,
we note that an exponential behavior of the resistivity is obtained only if the TO mode is sharp. If it is damped
(which inelastic neutron \cite{Yamada,Courtens1993}, THz \cite{misra2005}, and microwave \cite{marel_private} spectroscopies indicate), the exponential behavior is replaced by a power-law one; however, the exponent is still larger than 2 \cite{kumar_unpub}. 
Also, 
recent diagrammatic Monte Carlo calculations \cite{Mishchenko:2019} have shown that the onset of exponential behavior for a  Fr{\"o}hlich polaron is shifted down to lower temperatures due to mass renormalization; a similar effect can be expected for 2TO polarons.

Finally, we note that the 2TO model provides a falsifiable prediction because the scattering mechanism in this model is (quasi) elastic. This can be verified by checking if the electron part of the thermal conductivity and the electrical conductivity obey the Wiedemann-Franz law (if the model is valid, they should) and if the optical conductivity scales with $T/\omega$
(it should not).

We thank 
I. Aleiner,
C. Batista, 
K. Behnia,
P. Chandra,
A. Chubukov, C. Collignon, B. Fauqu{\'e}, 
C. Leighton, G. Lonzarich,  
A. Kreisel,
X. Lin,
S. Maiti, 
D. van der Marel,  E. Mikheev, I. Paul, 
P. Sharma, B. I. Shklovskii, and S. Stemmer for stimulating discussions,
and  K. Behnia and D. van der Marel for sharing their unpublished data with us. 
We acknowledge a financial support from the National Science Foundation under Grant No. DMR-1720816 (A.K. and D.L.M.), University of Florida under Opportunity Fund OR-DRPD-ROF2017 (A.K. and D.L.M.), and Basic Research Program of HSE (V.I.Yu.).

\bibliography{STOref_4}
\end{document}


\title{Supplemental Material for\\ ``Quasiparticle and Nonquasiparticle Transport in Doped Quantum Paraelectrics"}
\author{Abhishek Kumar}
\affiliation{University of Florida, Gainesville, Florida, 32611, USA}
 \author{Vladimir I. Yudson}
 \affiliation{Laboratory for Condensed Matter Physics, National Research University ``Higher School of Economics'', 
20 Myasnitskaya St., Moscow, 101000 Russia}
\affiliation{Russian Quantum Center, Skolkovo, Moscow 143025, Russia}
\author{Dmitrii L. Maslov}
\affiliation{University of Florida, Gainesville, Florida, 32611, USA}
\date{\today}
\begin{abstract}
\end{abstract}
\maketitle
\tableofcontents

\section{Electron self-energy for two-phonon scattering mechanism}
\label{sec:2ph}
Unlike single-phonon diagrams of the conventional diagrammatic technique, \cite{agd:1963} diagrams with two-phonon vertices acquire combinatorial coefficients arising from the number of ways  the correlation function $\langle \Pi_{n=1}^N P^2(x_n)\rangle$ (where $N$ is the number of two-phonon vertices and $x_n$ is the $n^{\text{th}}$ space-time point) can be partitioned into binary averages $\langle P(x_n) P(x_m)\rangle$. To find this number, we pick any particular $n$ and write $P^2(x_n)$ as $P(x_n) P(x_n)$. Now the first $P(x_n)$ can be paired with one of the two factors of $P^2(x_m)$ (with $m\neq n$), which gives a factor of 2. The remaining $P(x_n)$ van be paired with some other $P(x_k)$ also in two ways, hence another factor of two. This process continues until the last pairing, which can be done only in one way. Therefore, the combinatorial coefficient is equal to $2^{N-1}$. 
With our choice of the two-phonon coupling constant as in Eq.~(1) of the Main Text (MT), the $N^\text{th}$ order two-phonon diagram comes with a coefficient $(g_2/2)^N\times 2^{N-1}=g^N_2/2$ for any $N$, and there is no need to follow the combinatorial coefficients.

The two-loop two-phonon diagram for the electron self-energy is shown in Fig.~1a of the MT. 
Algebraically, 
\begin{widetext}
\bea
\label{selfenergy}
\Sigma(\bk, \epsilon_m) = 
\frac{1}{2}
T^2 \sum_{m_1,m_2} \int_{\bq_1, 
\bq} \sum_{\alpha\beta} 
\Gamma_{\alpha\beta}(\bq_1)\Gamma_{\beta\alpha}(\bq
-\bq_1) D(\bq_1, \omega_{m_1}) D(
\bq-\bq_1, \omega_{m_2}) G( \bk+
\bq, \epsilon_m+\omega_{m_1}+\omega_{m_2} ),\nn\\
\eea
\end{widetext}
where a factor of $1/2$ arises from the diagrammatic rule explained at the beginning of this section, $\int_\bp$ is a shorthand for $\int d^3p/(2\pi)^3$, $G
(\bp, \epsilon_m)=\left[i\e_m-\xi_\bp+\mu-\Sigma(\bp,\e_m)\right]^{-1}$ and $D(\bp,\omega_m)$ are the (Matsubara) electron and phonon Green's function respectively, $\xi_\bp=p^2/2m^*$ is the electron dispersion,
\bea
\label{ver}
\Gamma_{\alpha\beta}(
\bp)=g_2 \sum_{a=1,2} \frac{e^a_{\alpha\bp} e^a_{\beta
\bp}}{
4\pi} \left[ \ve_0(
\bp) - \ve_\infty \right] \omega_
\bp=g_2\left(\delta_{\alpha\beta} - \frac{
p_\alpha 
p_\beta}{
p^2}\right) \left[ \ve_0(
\bp) - \ve_\infty \right] \frac{\omega_
\bp}{4\pi}
\eea
is the electron-phonon interaction vertex, and $e^{a=1,2}_{\sigma={x,y,z}}(
\bp)$ are the components of the unit polarization vector of the $a^{\text{th}}$ branch of a TO mode with dispersion 
 \bea
 \omega^2_\bq=\omega_0^2+s^2q^2.\label{TOsp}
 \eea
In what follows, phonons will be treated as free bosons with a Green's function
\beq
\label{phmat}
\begin{split}
D_0 (
\bp, \omega_m) = -
2
 \frac{\omega_
\bp}{\omega_m^2 + \omega_
\bp^2}
\end{split}
\eeq
 For $T\gg\max\{\omega_0,T_{\text{BG}}\}$, phonons can be treated as static thermal disorder,  which corresponds to setting $\omega_m = 0$ in the phonon lines.
 (In Sec.~\ref{app:T2T5}, we showed that for STO this condition is reduced to $T\gg \omega_0$.) Next, we simplify the vertex by neglecting  $\ve_\infty$ compared to $\ve_0(\bq)$ in Eq.~\ref{ver} and eliminating $\ve_0(
 \bp)$ in favor of $\omega_
 \bp$ via the Lyddane-Sachs-Teller relation, $\ve_0(
 \bp) = \Omega_0^2/\omega_
 \bp^2$. This yields
 \bea
 \label{vertex}
 \Gamma_{\alpha\beta}(\bp)=g_2\left(\delta_{\alpha\beta} - \frac{
p_\alpha 
p_\beta}{
p^2}\right)\frac{\Omega_0^2}{4\pi \omega_\bp}.
 \eea

 We also assume for the time being that the material is very close to the quantum-critical point, so that $\omega_0$ can be set to zero, i.e.,  $\omega_\bp=sp$.    Summing over $\alpha,\beta$, we obtain an equation for the retarded self-energy $\Sigma(\bk,\epsilon)=\Sigma(\bk,i\e_m\to\e+i\delta)$
\beq
\label{se1}
\begin{split}
\Sigma(\bk, \epsilon) &= \frac{T^2 g_2^2 \Omega_0^2}{
8
\pi^2 s^4} \int_{\bq_1, \bq}G(\bk+\bq,\e)
 \frac{1}{q_1^2} \frac{1}{|\bq-\bq_
1|^2} \left( 1 + \frac{\left[\bq_1 \cdot (\bq-\bq_
1)\right]^2}{q_1^2 
|\bq-\bq_1|
^2} \right).
\end{split}
\eeq
The integral over $q_1$ is solved as 
\beq
\label{cal}
\begin{split}
& \int_0^{\infty} \frac{dq_1}{(2\pi)^2} q_1^2 \int_{-1}^1 {d(\cos\theta_1)}\int_0^{2\pi} \frac{d\phi}{2\pi} \frac{1}{q_1^2} \frac{1}{q^2 + q_1^2 - 2qq_1\cos\theta_1} \left( 1 + \frac{\left[ qq_1\cos\theta_1 - q_1^2 \right]^2}{q_1^2(q^2 + q_1^2 - 2qq_1\cos\theta_1)}  \right) \\
=& \frac{1}{4\pi^2 q} \int_0^\infty dy \int_{-1}^1 dx \frac{1}{1+y^2-2xy} \left( 1 + \frac{(x-y)^2}{1+y^2-2xy} \right)=\frac{3}{16q},
\end{split}
\eeq
where $x = \cos\theta_1$ and $y = q_1/q$. Then \bea\label{SU}
\Sigma(\bk, \epsilon) &=& \frac{3T^2 g_2^2 \Omega_0^2}{
128
\pi^2 s^4} \int_\bq G(\bk+\bq,\e)
\frac{1}{q}\equiv 
\int_\bq G(\bk+\bq,\e) U(\bq),
\eea
where
\bea
U(\bq)=\frac{3\pi T^2}{2 m^*q E_0}
\eea
with
\bea
\label{E0}
E_0=\frac{64\pi^3 s^4}{m^*g_2^2\Omega_0^4}
\eea
the characteristic energy scale of the model, defined in the MT. This reproduces Eqs.~(3) and (5) of the MT.

Relabeling $\bk'=\bk+\bq$ and integrating over the angle between $\bk$ and $\bk'$, we obtain an equivalent form of the Dyson equation for the self-energy, Eq.~(9) of the MT, which is reproduced here for reader's convenience:
\bea
\label{se3}
\Sigma(\xi, \epsilon)=\lambda \int_0^\infty d\xi' 
K(\xi,\xi')G(\xi',\e)
\eea
where  $\xi\equiv \xi_\bk$, $\xi'\equiv \xi_{\bk'}$, 
\bea
\label{lambda}
\lambda = 3T^2/4\pi E_0,\eea
and 
\bea
K(\xi,\xi')=\sqrt{\frac{\xi'}{\xi}} \Theta(\xi - \xi') + \Theta(\xi' - \xi).
\eea
Although the full solution of Eq.~(\ref{se3}) can be obtained only numerically (see Sec.~\ref{sec:num}), the asymptotic solutions in the quasiparticle ($T \ll E_0$) and non-quasiparticle ($T \gg E_0$) regimes can  be found analytically. This is the subject of Secs.~\ref{sec:qp} and \ref{sec:nqp}.

\subsection{Quasiparticle regime}
\label{sec:qp}
Suppose that  the self-energy is small and $G(\bk,\e)$ in Eq.~(\ref{SU}) can be replaced by its free-electron form $G_0(\bk,\e)=(\te-\xi_\bk+i0^+)$, where $\te=\e+\mu$. Then the quantum lifetime is related to the imaginary part of the self-energy on the mass-shell $\te=\xi_\bk$ via
\bea
\frac{1}{\tau_0}\equiv -2\im\Sigma(\bk,\e=\xi_\bk-\mu)=2\pi\int\frac{d^3q}{(2\pi)^3}\delta(\xi_\bk-\xi_{\bk+\bq})U(\bq)=\frac{1}{2\pi} \frac{m^*}{k}\int^{2k}_0 dqq U(\bq)\label{tau0}
\eea
(the last step is valid if $U(\bq)$ is isotropic, which is the case here). 
For short-range disorder ($U=\text{const}$), $1/\tau_0\propto k\propto \sqrt{\xi_\bk}$, which reflects the energy dependence of the density of states in 3D. But for our case of ``medium-range'' thermal disorder with $U\propto 1/q$, the factors of $k$ cancel each other and we obtain an energy-independent scattering rate
\bea
\label{sctt}
\frac{1}{\tau_0}=\frac{3}{2}
\frac{T^2}{E_0}.
\eea
The transport scattering rate differs from the quantum one by the presence of the $1-\cos\theta$ factor, which suppresses the contribution of small-angle scattering:
\bea
\frac{1}{\tau}=2\pi\int\frac{d^3q}{(2\pi)^3}\delta(\xi_\bk-\xi_{\bk+\bq})U(\bq)(1-\cos\theta).\label{tautr}
\eea
Since the integral in Eq.~(\ref{tau0}) is controlled by $q\sim k$, the transport factor in Eq.~(\ref{tautr}) modifies the result only by a number:
\bea
\label{tr_rate}
\frac{1}{\tau}=\frac{2}{3}\frac{1}{\tau_0}=\frac{T^2}{E_0}.
\eea
For elastic scattering, the Landau's condition \cite{physkin} for  quasiparticles to be well-defined reads 
\bea
\frac{1}{\tau}\ll\max\{T,E_F\}.\label{cond}
\eea
(Since $\tau\sim \tau_0$ in our case, it does not matter which of the two is  be used in the left-hand side of the inequality above.)
 In a typical experiment, $E_F\ll E_0$ and, therefore, Eq.~(\ref{cond}) is satisfied as long as $T\ll E_0$, regardless of the ratio of $T$ and $E_F$. If $E_F\gg E_0$,  the entire region of $T\ll E_0$ is within the degenerate regime, where Eq.~(\ref{cond}) is also satisfied. Therefore, the condition $T\ll E_0$ defines the quasiparticle regime of transport.

In what follows will need also the form of the self-energy for an arbitrary relation between $\te$ and $\xi$ (but still in the quasiparticle regime). The imaginary part of the self-energy can be readily obtained from Eq.~(\ref{se3}) by substituting $\im G(\xi',\e)=-\pi\delta(\tilde\e-\xi')$ in there and integrating over $\xi'$ with the result
\beq
\label{imsig_pert}
\text{Im}\Sigma(\xi, \epsilon) = -\pi\lambda \Theta(\tilde\epsilon) \left[ \sqrt{\frac{\tilde\epsilon}{\xi}} \Theta(\xi - \tilde\epsilon) + \Theta(\tilde\epsilon - \xi) \right].
\eeq
The $\Theta(\tilde\epsilon)$ factor in Eq.~\ref{imsig_pert} indicates that the $\text{Im}\Sigma$ is non-zero only above the bottom of the band (for $\tilde\epsilon>0$). 
For an on-shell particle with $\tilde\e=\xi$, $\text{Im}\Sigma(\xi, \xi-\mu)=-\pi \lambda=-3 T^2/4E_0$, which reproduces the result for $1/\tau_0$ in Eq.~(\ref{tau0}).

The real part of the self-energy can obtained from the Kramers-Kronig relation,
\beq
\label{KK}
\text{Re}\Sigma(\xi, \epsilon) = \frac{1}{\pi} P \int_
{-\infty}^\infty d\epsilon' \frac{\text{Im}\Sigma(\xi, \epsilon')}{\epsilon' - \epsilon},
\eeq
where $P$ denotes the principal value of the integral. From now onwards, we will denote $\text{Re}\Sigma \equiv \Sigma'$ and $\text{Im}\Sigma \equiv \Sigma''$. Substituting Eq.~(\ref{imsig_pert}) in Eq.~(\ref{KK}) and integrating over $\e'$, we obtain
\begin{subequations}
\bea
\label{rese1}
\Sigma'(\xi, \epsilon)& =& -\lambda \left[ P \int_0^\xi d\tilde\epsilon' \sqrt{\frac{\tilde\epsilon'}{\xi}} \frac{1}{\tilde\epsilon' - \tilde\epsilon} + P \int_\xi^\Lambda d\tilde\epsilon' \frac{1}{\tilde\epsilon' -\tilde \epsilon} \right]\label{Sp1}\\
&=& -\lambda \times
\left\{
\begin{array}{ccc}
2 - \sqrt{\frac{\te}{\xi}} ~ \text{ln}\frac{\sqrt{\xi} + \sqrt{\te}}{|\sqrt{\xi} - \sqrt{\te}|} + \text{ln} \frac{\Lambda}{|\xi - \te|},\;\text{for}\;\te>0,\\
2 - 2\sqrt{\frac{|\te|}{\xi}} \tan^{-1} \sqrt{\frac{\xi}{|\te|}} + \text{ln} \frac{\Lambda}{\xi + |\te|},\;\text{for}\;\te<0,
\end{array}
\right.
\label{Sp2}
\eea
\end{subequations}
where $\Lambda$ is the ultraviolet cutoff of the model (on the order of $1/m^*a_0^2$, which is the same as the electron bandwidth).
Note that here is no singularity in $\Sigma'$ on the mass-shell, i.e., at $\te=\xi>0$, because the log-divergent terms in the first line in the equation above cancel each other. However, the derivative of $\Sigma'$ diverges logarithmically near the mass-shell. Indeed, 
expanding the first line in Eq.~(\ref{Sp2}) near the mass-shell,  we get
\beq
\begin{split}
\Sigma'(\xi, \epsilon)
\approx -\lambda \left[ 2 + \text{ln} \frac{\Lambda}{\te} - 2 \text{ln} 2 
 +\frac{\te-\xi}{2\te} \left(1 - 2\text{ln}2 + \text{ln}\frac{|\xi-\te|}{\te}\right) \right]
\end{split}
\eeq
Even though the derivative of $\Sigma'$ diverges, the effective mass remains finite (and equal to the band mass) on the mass-shell. Indeed, the mass renormalization factor is given by
\beq
\frac{\tilde m}{m^*} = \frac{1 - \frac{\partial\Sigma'}{\partial\epsilon}|_{\te = \xi}}{1 + \frac{\partial\Sigma'}{\partial\xi}|_{\te = \xi}}.\label{mass_ratio}
\eeq
In our case, $\Sigma' = \lambda f(\xi/\te)$, the derivative with respect to $\te$ gives $(-\lambda\xi/\te^2) f'(\xi/\te)$. Similarly, the derivative of $\Sigma'$ with respect to $\xi$ gives $(\lambda/\te) f'(\xi/\te)$. Since $f' \propto \text{ln} |(\xi/\te)-1|$, the divergences in the numerator and denominator of Eq.~(\ref{mass_ratio}) cancel each other and $\tilde m/m^* = 1$.

\subsection{A differential form of the Dyson equation}
To analyze the non-quasiparticle regime, it is convenient to reduce the integral Dyson equation in Eq.~(\ref{se3}) to a differential form. To this end,  we  introduce two auxiliary self-energies, $\Sigma^<$ and $\Sigma^>$ as 
\beq
\Sigma = \Sigma^< + \Sigma^>,\label{sum}
\eeq
 where
\beq
\label{sig<1}
\Sigma^<(\xi, \epsilon) = \lambda \int_0^\xi d\xi' \sqrt{\frac{\xi'}{\xi}} \frac{1}{\te - \xi' - \Sigma(\xi', \epsilon)}
\eeq
and
\beq
\Sigma^>(\xi, \epsilon) = \lambda \int_\xi^\infty d\xi' \frac{1}{\te - \xi' - \Sigma(\xi', \epsilon)}.
\eeq
The auxiliary self-energies obey the differential equations:
\bse
\label{diff}
\beq
\frac{d}{d\xi} \left[ \sqrt{\xi} \Sigma^<(\xi, \epsilon) \right] = \lambda \sqrt{\xi} \frac{1}{\te - \xi
 - \Sigma(\xi
 , \epsilon)},\label{<3}
\eeq
\beq
\frac{d}{d\xi}\Sigma^
>(\xi, \epsilon) = -\lambda \frac{1}{\te - \xi
 - \Sigma(\xi
 , \epsilon)}.\label{>3}
\eeq
\ese
A corollary of  Eqs.~(\ref{<3}) and (\ref{>3}) is another differential equation:
\beq
\label{de1}
\frac{1}{\sqrt{\xi}} \frac{d}{d\xi} \left[ \sqrt{\xi} \Sigma^<(\xi, \epsilon)  \right] + \frac{d}{d\xi} \Sigma^>(\xi, \epsilon) = 0.
\eeq
Opening the derivative in the equation above and recalling Eq.~(\ref{sum}), we arrive at yet another equation
\beq
\label{sig<2}
\Sigma^<(\xi, \epsilon) = -2\xi \frac{d}{d\xi} \Sigma(\xi, \epsilon).
\eeq
Multiplying the last equation by $\sqrt{\xi}$, differentiating the result over $\xi$, and using 
Eq.~(\ref{<3}), we obtain 
a non-linear differential equation for $\Sigma(\xi, \epsilon)$: 
\beq
\label{non-linear}
\frac{1}{\sqrt{\xi}} \frac{d}{d\xi} \left[ \xi^{3/2} \frac{d}{d\xi} \Sigma(\xi, \epsilon) \right] = -\frac{\lambda}{2} \frac{1}{\te - \xi- \Sigma(\xi, \epsilon)}.
\eeq

\subsection{Self-energy in the non-quasiparticle regime}
\label{sec:nqp}
\subsubsection{Linearized differential Dyson equation}
\label{sec:linear}
We now turn to the non-quasiparticle regime of $T \gg E_0$. 
In the quasiparticle regime, the spectral function $G''=\Sigma''/\left[(\te-\xi-\Sigma')^2+(\Sigma'')^2\right]$ is non-zero only for $\te>0$, i.e., for particles with energies above the bottom of the band. We will show at strong coupling the threshold in the self-energy is shifted to a finite energy, which depends on $\lambda$. 
In the MT, it was shown that the $\re\Sigma$ depends on $\e$ and $\xi$ only logarithmically. Neglecting this weak dependence, we approximate $\re\Sigma$ by a constant and absorb this constant into the chemical potential. 
Taking the imaginary part of Eq.~(\ref{non-linear}) yields:
\beq
\label{sig2_nl}
\frac{1}{\sqrt{\xi}} \frac{d}{d\xi} \left[ \xi^{3/2} \frac{d}{d\xi} \Sigma''(\xi, \epsilon) \right] = -\frac{\lambda}{2} \frac{\Sigma''(\xi, \epsilon)}{\left[ \te - \xi \right]^2 + \Sigma''^2(\xi, \epsilon)}.
\eeq
Suppose that a threshold in $\e$ does exist.  If we also assume (to be verified later) that $\e+\mu<0$ near the threshold, while $\xi>0$ by definition, then there is no quasiparticle pole, i.e., the $(\dots)^2$ term in the denominator of the RHS does not vanish. On the other hand, $\Sigma''(\xi, \epsilon)$ vanishes at the threshold and must be small right above the threshold. Then the RHS of Eq.~(\ref{sig2_nl}) can be expanded in $\Sigma''(\xi, \epsilon)$. Introducing $\bar{\epsilon} = -\te$ (by assumption, relevant $\bar\e>0$), denoting $\gamma_\e(\xi)\equiv-\Sigma''(\xi, \epsilon)$ and retaining only the leading in $\gamma_\e(\xi)$ term on the RHS of Eq.~(\ref{sig2_nl}), we obtain a linear equation
\beq
\frac{1}{\sqrt{\xi}} \frac{d}{d\xi} \left[ \xi^{3/2} \frac{d}{d\xi} \gamma_{\e}(\xi) \right] = -\frac{\lambda}{2} \frac{\gamma_{\e}(\xi)}{\left( \bar\e +\xi\right)^2}.\label{gamma_diff}
\eeq
The threshold $\e_0$ is the largest value of $\bar\e$ at which this equation still has a non-trivial solution.

We now introduce a new function via 
\bea
\gamma_{\e}(\xi) = \xi^a \varphi_{\e}(\xi),
\eea 
and choose $a$ in such a way that the $d\varphi_{\e}/d\xi$ term is eliminated. This gives $a=-3/4$ and we obtain 
\beq
\label{susy_ham}
-\varphi''_{\e}(\xi) - \left[ \frac{3}{16\xi^2} + \frac{\lambda}{2\xi (\bar{\epsilon} + \xi)^2} \right] \varphi_{\e}(\xi) = 0,
\eeq
which is Eq.~(11) of the MT. Equation (\ref{susy_ham}) can be solved in two ways: straightforwardly and via supersymmetric quantum mechanics.

\subsubsection{Straightforward solution of Eq.~(\ref{susy_ham})}
\label{sec:simple}

A general solution of Eq.~(\ref{susy_ham}) can be written as
\beq
\begin{split}
\varphi_{\e}(\xi) = \text{const} \times \xi^{1/4} (\xi+\bar{\epsilon})^{1/2} \times \begin{cases} &\cos\Omega(\xi) \\
&\sin\Omega(\xi),
\end{cases}
\end{split}
\eeq
or
\beq
\begin{split}
\gamma_{\e}(\xi) = \text{const} \times \sqrt{\frac{\xi+\bar\e}{\xi}} \times \begin{cases} &\cos\Omega(\xi) \\
&\sin\Omega(\xi),
\end{cases}
\end{split}
\eeq
where 
\beq
\Omega(\xi) = \sqrt{1+\frac{2\lambda}{\bar{\epsilon}}} \tan^{-1} \sqrt{\frac{\xi}{\bar{\epsilon}}}.
\eeq
The integral
Dyson equation,
Eq.~(\ref{se3}),
as well as the physical content of the problem, impose two boundary conditions: $\gamma_{\e}(0) < \infty$ and $\gamma_{\e}(\xi \to \infty) \to 0$. The solution that satisfies the first condition corresponds to a choice
\beq
\gamma_\e(\xi) = \text{const} \times \sqrt{\frac{\xi+\bar{\epsilon}}{\xi}} \sin\Omega(\xi).\label{sine}
\eeq
At $\xi \to \infty$, however, the ``frequency" $\Omega(\xi)$ approaches a finite value of $\Omega(\infty) = (\pi/2) \sqrt{1+2\lambda/\bar{\epsilon}}$, whereas the prefactor of the sine function in Eq.~(\ref{sine}) approaches unity. Therefore, $\gamma(\xi)$ vanishes at $\xi \to \infty$ only if $\Omega(\infty) = \pi n$, which implies that $\bar{\epsilon} = 2\lambda/(4n^2 - 1)$. The $n=1$ eigenvalue $\bar\e=2\lambda/3$ is the largest value of $\bar\e$ at  which Eq.~(\ref{susy_ham}) still has a non-trivial solution.\footnote{A similar situation occurs in the theory of superconductivity near a quantum critical point, see Ref.~\onlinecite{abanov2020}.} This means the threshold is located at 
\bea
\e_0=2\lambda/3=\frac{T^2}{2\pi E_0}.\label{bare0}
\eea Right at the threshold, the oscillatory solution in Eq.~(\ref{sine}) is reduced to a rational form: 
\beq
\label{gamma0}
\gamma_{\e_0}(\xi) = \text{const} \times \sqrt{\frac{\xi + \e_0}{\xi}} \sin \left[ \sqrt{1 + \frac{2\lambda}{\e_0}} \tan^{-1} \sqrt{\frac{\xi}{\e_0}} \right] = \frac{\text{const} }{\sqrt{\xi + \e_0}}.
\eeq

\subsubsection{Supersymmetric solution of Eq.~(\ref{susy_ham})}
\label{sec:susy}
The existence of a threshold has an interesting interpretation in terms of supersymmetric (SUSY) quantum mechanics. Indeed, Eq.~(\ref{susy_ham}) can be viewed as a Schroedinger equation for the zero-energy eigenstate.  Such an eigenstate always exists if the corresponding Hamiltonian, $H = -\partial^2_\xi + V(\xi)$, with
\beq
\label{susy_pot}
V(\xi) = -\frac{3}{16\xi^2} - \frac{\lambda}{2\xi(\xi+\bar{\epsilon})^2},
\eeq
is supersymmetric.  For this to be the case, the Hamiltonian must be of the form $H=Q^\dagger Q$ (see, e.g., Ref.~\onlinecite{Cooper:1995}), where  
\beq
Q = \partial_\xi + W(\xi) \,\,\,\,\,\, \text{and} \,\,\,\,\,\, Q^\dagger = -\partial_\xi + W(\xi).
\eeq
Equating the original and SUSY forms of $H$, we arrive at a non-linear Riccati equation for the superpotential, $W(\xi)$:
from where
\beq
\label{riccati}
W'(\xi) - W^2(\xi) = -V(\xi) = \frac{3}{16\xi^2} + \frac{\lambda}{2\bar{\epsilon}^2} \left[ \frac{1}{\xi} - \frac{1}{\xi+\bar{\epsilon}} \right] - \frac{\lambda}{2\bar{\epsilon}} \frac{1}{(\xi+\bar{\epsilon})^2}.
\eeq
We search for $W(\xi)$ in the following form
\beq
W(\xi) = \frac{A}{\xi} + \frac{B}{\xi+\bar{\epsilon}},
\eeq
where $A$ and $B$ are constants.
Then
\beq
\label{W}
W'(\xi) - W^2(\xi) = -\frac{A+A^2}{\xi^2} - \frac{B+B^2}{(\xi+\bar{\epsilon})^2} - \frac{2AB}{\bar{\epsilon}} \left[ \frac{1}{\xi} - \frac{1}{\xi+\bar{\epsilon}} \right].
\eeq
Comparing similar terms in Eqs.~\ref{riccati} and (\ref{W}), we find a system of equations for the coefficients $A$ and $B$:
\bse
\beq
\label{s1}
A^2 + A + \frac{3}{16} = 0,
\eeq
\beq
\label{s2}
B^2 + B - \frac{\lambda}{2\bar{\epsilon}} = 0,
\eeq
\beq
\label{s3}
AB = -\frac{\lambda}{4\bar{\epsilon}}.
\eeq
\ese
From Eq.~(\ref{s1}), we find $A_+ = -1/4$ and $A_- = -3/4$. Introducing $r = \lambda/\bar{\epsilon}$, we find from Eq.~(\ref{s3}) that $B_+ = -r/(4A_+) = r$ and $B_- = -r/(4A_-) = r/3$, respectively. Then Eq.~(\ref{s2}) yields either
\beq
r^2 + r - r/2 = 0 \to r = -1/2,
\eeq
or
\beq
\frac{1}{9}r^2 + \frac{1}{3}r - \frac{1}{2}r = 0 \to r = 3/2.
\eeq
Since $r$ must be non-negative, only the second solution makes sense. This corresponds to $\bar{\epsilon}/\lambda=1/r = 2/3$. Accordingly, the superpotential becomes $W(\xi) = -3/4\xi + 1/2(\xi+\bar{\epsilon})$. The first condition implies that the threshold in the imaginary part of the self energy is at $\e_0 = 2\lambda/3$. The zero-energy eigenstate of $H$ is readily found from the first-order supersymmetric equation, $Q\varphi_{\e_0}(\xi) = 0$, which yields $\varphi_{\e_0}(\xi) = \text{const}\times \xi^{3/4}/\sqrt{\xi+\e_0}$ or  $\gamma_{\e_0}(\xi) = \text{const}/\sqrt{\xi+\e_0}$. These are the same results as obtained in Sec.~\ref{sec:simple}.

\subsubsection{Self-energy above the threshold}
In Secs.~\ref{sec:simple} and \ref{sec:susy} we showed that the imaginary part of the self-energy vanishes identically below the threshold. Now we need to find its form slightly above the threshold. To this end, we go back to the integral form of the Dyson equation, Eq.~(\ref{se3}), and take its imaginary part. In the notations of Sec.~\ref{sec:linear}, we have :
\beq
\label{thresh}
\gamma_{\e}(\xi) = \lambda \int_0^\infty d\xi' 
K(\xi,\xi')\frac{\gamma_{\e}(\xi')}{(\bar{\epsilon} + \xi')^2 + \gamma^2_{\e}(\xi')}.
\eeq
Following the same logic as in the previous Section, we again expand the RHS of Eq.~(\ref{thresh}) in $\gamma_{\e}$ but now  keep  the cubic term: 
\beq
\label{thresh1}
\gamma_{\e}(\xi) = \lambda \int_0^\infty d\xi' 
K(\xi,\xi')\left[ \frac{\gamma_{\e}(\xi')}{(\bar{\epsilon} + \xi')^2} - \frac{\gamma^3_{\e}(\xi')}{(\bar{\epsilon} + \xi')^4} \right].
\eeq
The linearized version of this equation (without the cubic term on its RHS) is equivalent to the differential equation (\ref{gamma_diff}), whose solution right at the threshold is given by Eq.~(\ref{gamma0}). Now we search for a solution of Eq.~(\ref{thresh1}) close to but finite distance from the threshold in the following form
 \bea
 \gamma_{\e}(\xi) = C(\bar{\epsilon})/\sqrt{\xi+\e_0},
 \label{gamma1}
 \eea 
 assuming that $\bar{\epsilon} =\epsilon_0 - \delta\epsilon$ with $\delta\e\ll\e_0$. When substituting  Eq.~(\ref{gamma1}) into Eq.~(\ref{thresh1}),   we expand the first term  on the RHS of Eq.~(\ref{thresh1}) to first order in $\delta\e$  and set $\delta\e=0$ in the second term.
  This yields:
\beq
\label{thresh2}
\frac{C(\bar{\epsilon})}{\sqrt{\xi + \e_0}} = \lambda \int_0^\infty d\xi' K(\xi, \xi') \left[ \frac{C(\bar{\epsilon})}{(\xi'+\e_0)^{5/2}} + \frac{2 \delta\epsilon C(\bar{\epsilon})}{(\xi'+\e_0)^{7/2}} - \frac{C^3(\bar{\epsilon})}{(\xi'+\e_0)^{11/2}} \right],
\eeq
The first term on the RHS of Eq.~(\ref{thresh2}) cancels out with the left-hand side, and from the remaining terms we find  $C(\bar{\epsilon})$ as \beq
C^2(\bar{\epsilon}) = 
2\delta\epsilon \frac{\int_0^\infty d\xi' K(\xi, \xi') (\xi' + \e_0)^{-7/2}}{\int_0^\infty d\xi' K(\xi, \xi') (\xi' +\e_0)^{-11/2}} = 
\delta\epsilon
\e_0
 (\xi+\e_0) 
S^2\left({\xi}/{\e_0}\right),
\eeq
where
\beq
\label{S}
S(x) = \left[ \frac{42 (x+1) (2x+3)}{16x^3 + 56x^2 + 70x + 35} \right]^{1/2}.
\eeq 
The result for $\gamma_{\e}(\xi)$ reads
\beq
\label{gamma}
\gamma_{\e}(\xi) = \sqrt{\left(\e_0-\be
\right) \e_0}\times S(\xi/\e_0)= \sqrt{\left(\e_0+\e+\mu
\right) \e_0}\times S(\xi/\e_0).
\eeq
It can be readily checked that  $S(0) = 3\sqrt{2/5}$ and  $S(x\to\infty) = \sqrt{21/4x}$. Therefore, $\gamma_{\e}(0)$ is indeed finite and $\gamma_{\e}(\xi\to\infty)\to 0$, in agreement with the boundary conditions imposed when solving Eq.~(\ref{gamma_diff}) by the method of Sec.~\ref{sec:simple}.
\subsubsection{Position of the chemical potential in the non-quasiparticle regime}
\label{sec:mu}
The chemical potential is found from the condition of fixed number density:
\bea
\label{n}
n&=&\int_{-\infty}^\infty \frac{d\e}{\pi} n_F(\e) \int_0^\infty d\xi N(\xi) \xi (-)G''(\xi, \epsilon),\nn\\
&=&\frac{\sqrt{2} {m^*}^{3/2}}{\pi^2} \int_{-\infty}^\infty d\epsilon n_F(\e)\int_0^\infty d\xi \sqrt{\xi} \frac{\gamma_{\e}(\xi)}{\left( \epsilon-\xi+\mu \right)^2 + \gamma_{\e}^2(\xi)},
\eea
where $N(\xi)=(m^*)^{3/2}\sqrt{2\xi}/\pi^2$ is the density of states in 3D and where we again absorbed $\Sigma'$ into $\mu$. The non-interacting electron gas would be non-degenerate for $T\gg E_F$. We assume (and later verify) that this is also true even in the non-quasiparticle regime; accordingly, $n_F(\e)=e^{-\e/T}$.  Introducing again $\bar\e=-(\e+\mu)$ and recalling that $\gamma_{\e}(\xi)=0$ for $\e<\e_0=2\lambda/3\gg  T$, we obtain
\bea
n = \frac{\sqrt{2} {m^*}^{3/2}}{\pi^2} e^{\mu/T}\int_{-\infty}^{\e_0} d\bar\epsilon e^{\bar\e/T}\int_0^\infty d\xi \sqrt{\xi} \frac{\gamma_{\e}(\xi)}{\left(\bar\epsilon+\xi \right)^2 + \gamma_{\e}^2(\xi)}.
\eea
Because of the Boltzmann factor in the integrand, the integral is controlled by the region $\bar\e\approx\e_0$. In this region, we can neglect $\gamma_{\e}(\xi)$ in the denominator and replace $\gamma_{\e}(\xi)$ in the numerator by Eq.~(\ref{gamma}). Also, the lower limit of the $\bar\e$ integral can be replaced by $0$ and $\bar\e$ in the remaining term in the denominator can be replaced by $\e_0$. After these steps,
\bea
n &=& \frac{\sqrt{2} {m^*}^{3/2}}{\pi^2} e^{\mu/T}\int_{0}^{\e_0} d\bar\epsilon e^{\bar\e/T}\sqrt{(\e_0-\bar\e)\e_0}\int_0^\infty d\xi \frac{\sqrt{\xi} S(\xi/\e_0)}{\left(\bar\epsilon+\xi \right)^2}\nn\\
&=& \frac{c_0\sqrt{2} {m^*}^{3/2}}{\pi^2} e^{\mu/T}\int_{0}^{\e_0} d\bar\epsilon e^{\bar\e/T}\sqrt{\e_0-\bar\e}=\frac{c_0\sqrt{2} {m^*}^{3/2}}{2\pi^{3/2}} e^{\mu/T} e^{\e_0/T}T^{3/2},\label{n2}
\eea
where at the last step we used that $T\ll\e_0$ and 
\bea
c_0=\int^\infty_0 \frac{\sqrt{x}S(x)}{(x+1)^2}={1.46\dots}
\eea
Solving for $\mu$ and expressing $n$ via the Fermi energy, we obtain
\bea
\mu=-\e_0-\frac{3}{2}T\ln\frac{T}{E_F}+\frac{8}{3}T.
\eea
To leading log order, the last term can be neglected, and we arrive at the result quoted in the MT. The second term,  which coincides with the chemical potential of a free Boltzmann gas, is smaller in magnitude than the first term resulting from the interaction with phonons, as long as $T\gg E_0$, i.e., in the non-quasiparticle regime. By the same condition, the first term is larger in magnitude that $T$, and we are indeed in the non-degenerate regime. 
Now we see that the threshold in the original energy $\e$ is at
\bea
\epsilon_1=-\e_0-\mu=\frac{3}{2}T\ln\frac{T}{E_F}-\frac{8}{3}T\ll \e_0.
\eea

\subsection{Numerical solution of the Dyson equation}
\label{sec:num}
In this section, we provide details of the numerical self-consistent solution of the Dyson equation, Eq.~(\ref{se3}). We rescale all the energies by $E_0$ and write down  real and imaginary parts of Eq.~(\ref{se3}) as
\bse
\beq
\label{sig1}
\Sigma_1(E, w) = \frac{3}{4\pi} t^2 \int_0^\infty dE' \left[ \sqrt{\frac{E'}{E}} \Theta(E-E') + \Theta(E'-E) \right] \frac{w - E' - \Sigma_1(E', w)}{\left[ w - E' - \Sigma_1(E', w) \right]^2 + \Sigma_2^2(E', w)}
\eeq
and
\beq
\label{sig2}
\Sigma_2(E, w) = \frac{3}{4\pi} t^2 \int_0^\infty dE' \left[ \sqrt{\frac{E'}{E}} \Theta(E-E') + \Theta(E'-E) \right] \frac{\Sigma_2(E', w)}{\left[ w - E' - \Sigma_1(E', w) \right]^2 + \Sigma_2^2(E', w)},
\eeq
\ese
where $\Sigma_1 = \Sigma'/E_0$, $\Sigma_2 = -\Sigma''/E_0=\gamma_\e(\xi)/E_0$, $E = \xi/E_0$, $E'=\xi'/\e_0$, $w = \epsilon/E_0$ and $t = T/E_0$. 
Contrary to what was done in the previous sections, where $\Sigma'$ was assumed to be constant and absorbed into the chemical potential, here we do the opposite--the chemical potential will be absorbed into $\Sigma'$. We will not attempt to find the chemical potential self-consistently, as its position does not affect the resistivity, calculated in Sec.~\ref{resistivity}. To solve Eqs.~(\ref{sig1}) and (\ref{sig2}) self-consistently, we need to provide initial guesses for $\Sigma_1(E, w)$ and $\Sigma_2(E, w)$ and run the code  until the solution converges. As an initial guess of $\Sigma_1$, we use the perturbative solution given by Eqs.~(\ref{Sp1}) and (\ref{Sp2}) (with all the energies rescaled by $E_0$). On the other hand, the perturbative solution in Eq.~(\ref{imsig_pert}) is a bad initial guess for $\Sigma_2$ because  it is zero for $w<0$, while we need to allow for finite spectral weight at $w<0$.
This is achieved by replacing $\Theta(w)$ in Eq.~(\ref{imsig_pert}) by a smooth function with a non-zero support at $w<0$, e.g., $(1+\tanh w)/2$. The initial guesses, therefore, are chosen as 
\bse
\beq
\Sigma_1(E, w) = - \frac{3}{4\pi} t^2 \left[ \Theta(-w) \left\{ 2 - 2\sqrt{\frac{|w|}{E}} \tan^{-1} \sqrt{\frac{E}{|w|}} + \text{ln} \frac{L}{E+|w|} \right\} + \Theta(w) \left\{ 2 - \sqrt{\frac{w}{E}} \text{ln} \frac{\sqrt{E} + \sqrt{w}}{|\sqrt{E} - \sqrt{w}|} + \text{ln}\frac{L}{|E-w|} \right\} \right]
\eeq
and
\beq
\Sigma_2(E, w) = \frac{3t^2}{8} \left[ 1 + \tanh w \right] \left[ \sqrt{\frac{w}{E}} \Theta(E-w) + \Theta(w-E) \right],
\eeq
\ese
where $L=\Lambda/E_0$. 

We perform  numerical integration in Eqs.~(\ref{sig1}) and (\ref{sig2}) using the standard Riemann sum over a regular partition of the integration variable $E$. The step size and the upper limit for $E$-integral are chosen to be $\Delta E = 0.04$ and $E_{\max} = 837$ 
(this corresponds to $3\Lambda/8\pi E_0=100$, which is more ``natural'' choice of units).  The energy ($w$) is an external parameter varied in steps $\Delta w$ from 0.078 to 0.66, depending on the temperature ($t$).  The change in the step size is dictated by the subsequent need to integrate over $w$ in the Kubo formula for the resistivity, Eq.~(\ref{res}).
For small $t$, we choose $\Delta w$ and the upper limit $w_{\max}$ as small as possible in order to accurately capture the position of the threshold as well ensure optimal convergence of the $w$-integral in Eq.~(\ref{res}). On the other hand, since larger $t$ require also larger $w_{\max}$, we choose larger $\Delta w$ to reduce the computation time. For example, for $t = 0.42$ (the lowest $t$ value in our calculation), we choose $\Delta w = 0.078$ and $w_{\max} = 8.37$. 
For $t = 25.12$ (the highest value of $t$ in our calculation), we choose  $\Delta w$ and $w_{max}$ are $0.66$ and $167.47$, correspondingly. 
For small $t$, it usually takes roughly 15-20 iterations to converge, while for higher $t$  the number of iterations goes up to 50. The solution is not sensitive to the choice of the initial guess, as long as the former is a regular function.

\begin{figure}
\centering
\includegraphics[scale=0.4]{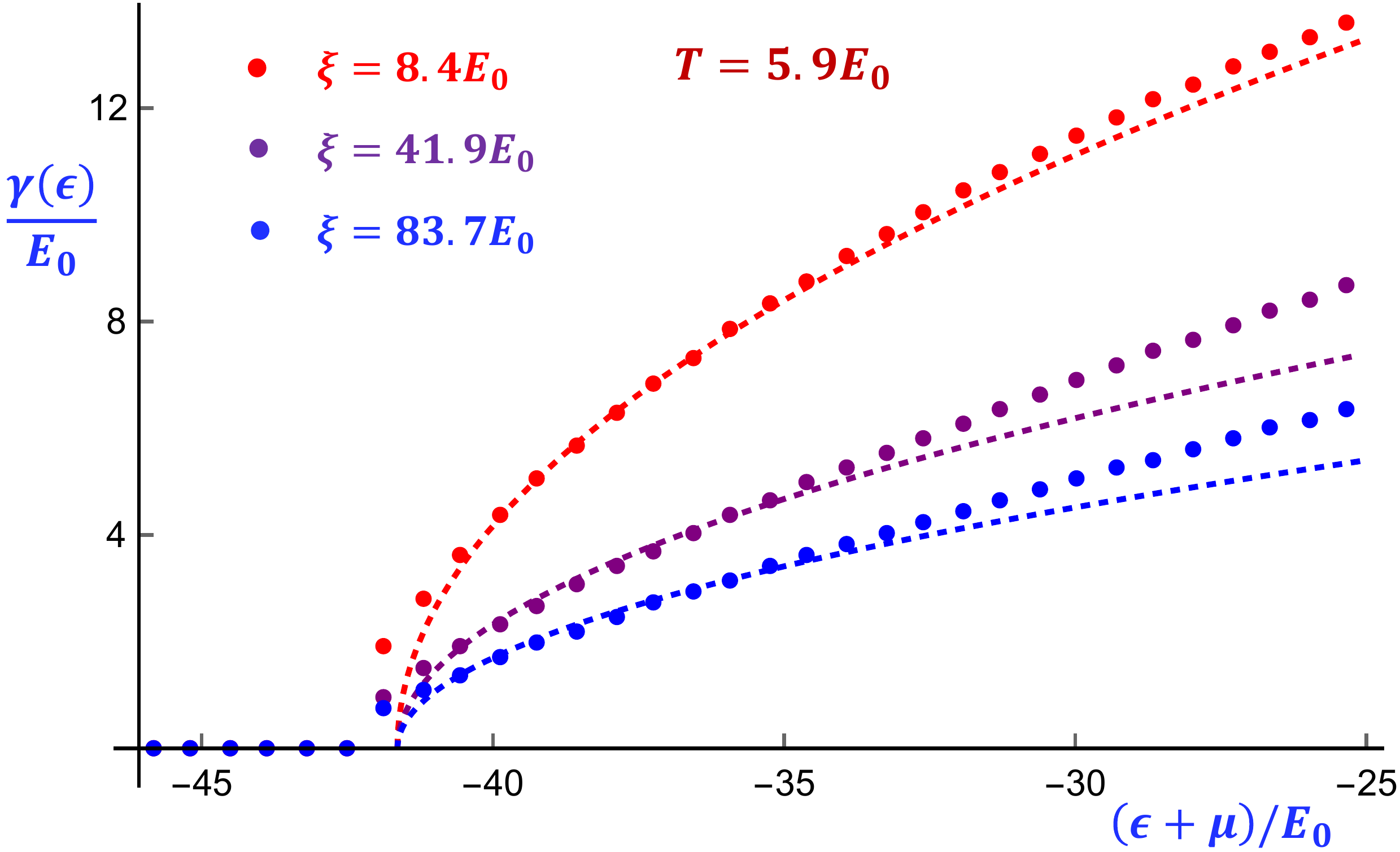}
\caption{\label{ImSig_omega} The threshold behavior of the imaginary part of the self-energy as a function of $\e+\mu$ for a range of electron dispersions, $\xi$, as specified in the legend, and for $T=5.9E_0$, i.e., in the non-quasiparticle regime.  Dots: numerical results; dashed lines:  analytical result in Eq.~(\ref{gamma}). Here, $\gamma_\e(\xi)=-\Sigma''(\xi,\e)$ and $E_0$ is defined in Eq.~(\ref{E0}).}
\end{figure}

Figure \ref{ImSig_omega} shows numerically calculated $
\gamma_\e(\xi)=-\Sigma''(\xi,\e)$ in the threshold region (dots) at $t \equiv T/E_0 = 5.9$  for a range of $E \equiv \xi/E_0$, as specified in the legend. 
The dashed lines are obtained from the analytic result  in Eq.~(\ref{gamma}). We remind the reader that the threshold position is determined up to the chemical potential; therefore, the horizontal axis corresponds to $\e+\mu$. 
In Fig.~\ref{fit_thr}, we show the dependence of the threshold in the variable $\bar\e=-(\e+\mu)$ as a function of $T$. In Sec.~\ref{sec:nqp}, we found that if the real part of the self-energy is assumed to be constant, then the threshold in $\bar\e$  is at $\e_0=2\lambda/3$. Here, we find that the actual threshold in $\e+\mu$ is fitted better by the function \bea
\label{fit_thr}
\tilde\e_0=a_1 \lambda\ln\left(\frac{a_2\Lambda}{\lambda}\right),
\eea
with 
$a_1=0.88$
 and $a_2=3.4$.
Given that the real part self-energy depends on $\e$ logarithmically rather than being constant, the difference between the linear dependence of the analytic result on $\lambda$ and the $\lambda\ln\lambda$ dependence of the numerical result is understandable.
  
\begin{figure}
\centering
\includegraphics[scale=0.4]{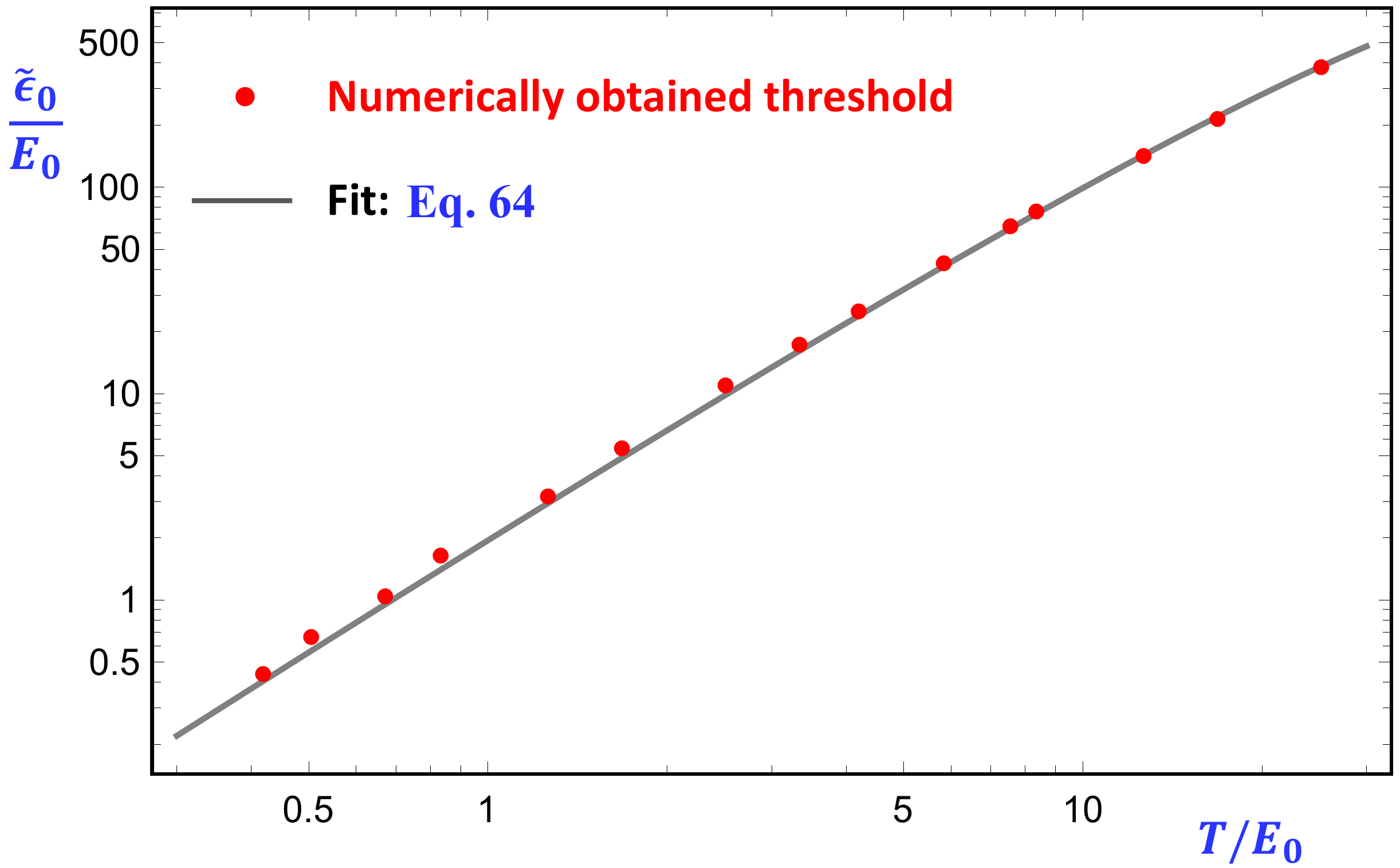}
\caption{\label{pos_thr} Position of the threshold as a function of the temperature. Points: numerical results. Solid line: fit by Eq.~(\ref{fit_thr}) with $a_1=0.88$ and $a_2=3.4$.}
\end{figure}
\label{shift_thr}

\section{Resistivity in the quasiparticle and non-quasiparticle regimes}
\label{resistivity}
\subsection{Kubo formula}
\label{sec:kubo}
Neglecting the vertex correction of both ladder and Cooperon types, the Kubo formula for the conductivity of electrons with parabolic spectrum reads
\bea
\label{sigma}
\sigma=\frac{2e^2}{3m^*}\int_{-\infty}^\infty \frac{d\epsilon}{\pi}\left(-\frac{\partial n_F(\e)}{\partial \e}\right) \int_0^\infty d\xi N(\xi) \xi \left[G''(\xi, \epsilon) \right]^2,
\eea
where $N(\xi)=m^{*3/2}\sqrt{2\xi}$ is the density of states in 3D. 
The number density $n$ is related to the spectral function via the first line in Eq.~(\ref{n}).
 It will be convenient to multiply and divide  Eq.~(\ref{sigma}) by carrier number density, $n$. Then resistivity is given by
\bea
\rho=\sigma^{-1}= \frac{3m^*}{2ne^2}\frac{\int_{-\infty}^\infty d\te 
n_F(\te)
 \int_0^\infty d\xi N(\xi) (-) G''(\xi, \te)}{\int_{-\infty}^\infty d\te \left(-\frac{\partial n_F(\te)}{\partial \te}\right)\int_0^\infty d\xi N(\xi) \xi \left[G''(\xi, \te) \right]^2},
\eea 
where $\te=\e+\mu$. 
In the non-degenerate regime, $n_F(\te)=e^{(\mu-\te)/T}$ while $-n'_F(\te)=e^{(\mu-\te)/T}/T$, and thus the $e^{\mu/T}$ factor cancels out between the numerator and denominator in the equation above.  Then the resistivity is reduced to
\bea
\label{res}
\rho=\sigma^{-1}= \frac{3m^*T}{2ne^2}\frac{\int_{-\infty}^\infty d\te 
e^{-\te/T}
 \int_0^\infty d\xi N(\xi) (-) \text{Im}G(\xi, \te)}{\int_{-\infty}^\infty d\te e^{-\te/T} \int_0^\infty d\xi N(\xi) \xi \left[ \text{Im}G(\xi, \te) \right]^2},\label{rho2}
\eea
which is Eq.~(15) of the MT. 
\subsection{Quasiparticle regime}
In the quasiparticle regime, the spectral function in the numerator of Eq.~(\ref{rho2}) can be replaced by its free-electron form. Then the integral in numerator is reduced to
\beq
\int_{-\infty}^\infty d\te e^{-\te/T} \int_0^\infty d\xi \sqrt{\xi} (-)G''(\xi,\e)=\pi\int_{-\infty}^\infty d\te e^{-\te/T} \int_0^\infty d\xi \sqrt{\xi} \delta(\te-\xi) = \frac{\pi^{3/2}}{2}T^{3/2}.
\eeq
The integral in denominator is  simplified by replacing the imaginary part of the self-energy by its perturbative form, Eq.~(\ref{imsig_pert}) and neglecting the real part of the self-energy, which is justified  to leading order in the coupling constant, $\lambda$.  Then
\bea
&&\int_{-\infty}^\infty d\te e^{-\te/T} \int_0^\infty d\xi \xi^{3/2} \left[G''(\xi,\te)\right]^2=\int_{0}^\infty d e^{-\te/T} \int_0^\infty d\xi \xi^{3/2} \left(\frac{\Sigma''(\xi,\te)}{(\te-\xi)^2+\left[\Sigma''(\xi,\te)\right]^2}\right)^2\nn\\
&\approx &\int_{0}^\infty d\te e^{-\te/T}\te^{3/2}\left[\Sigma''(\te,\te)\right]^2 \int_{-\infty}^\infty d\bar\xi \frac{1}{\left(\bar\xi^2+\left[\Sigma''(\te,\te)\right]^2\right)^2}=\frac{\pi}{2} \int_{0}^\infty d\te e^{-\te/T}\e^{3/2}\frac{1}{|\Sigma''(\te,\te)|}
=\frac{3\sqrt{\pi}T^{5/2}}{8\lambda},
\eea
where $\bar\xi=\xi-\te$. Using Eq.~(\ref{lambda}) for $\lambda$, we find
\beq
\label{res_lowT}
\rho = \frac{3 m^* T^2}{2 n e^2 E_0}.
\eeq
The ``transport correction'', resulting from resumming ladder diagrams, eliminates the factor of $3/2$, and the resistivity coincides with the expression given in Eq.~(8)  of the MT.

\subsection{Non-quasiparticle regime}
As in the calculation of the chemical potential (Sec.~\ref{sec:mu}), the integrals over $\te$ in Eq.~(\ref{rho2}) are controlled by a narrow region near the threshold. Repeating the same steps as in Eqs.~(\ref{n}-\ref{n2}), we arrive at

\beq
\label{res_largeT}
\rho = C \frac{m^*}{ne^2} \sqrt{T\e_0} \propto T^{3/2}
\eeq
where we used Eq.~(\ref{bare0}) for $\e_0$ and
\bea
\label{C}
C=\frac{3\sqrt{\pi}}{4} \frac{\int^\infty_0 dx \frac{x^{1/2}}{(x+1)^{2}} S(x)}{\int^\infty_0 dx \frac{x^{3/2}}{(x+1)^{4}}S^2(x)}\approx 5.6.\label{Crho}
\eea
The numerical solution along with the analytic results in quasiparticle and non-quasiparticle regimes, Eqs.~(\ref{res_lowT}) and (\ref{res_largeT}) correspondingly, are shown in the inset of Fig.~2 of the MT.

\section{Higher-order processes}

 In this section, we  discuss the role of higher-order diagrams, shown in Figs.~$1b-1f$ of the MT. We will demonstrate that the non-crossing diagrams forming an ``umbrella'' sequence, i..e, diagrams 1{\em a}-1{\em c} and other diagrams of the same type, can be resummed into a closed form. The result for the two-loop diagram (Fig. 1{\em a}) is reproduced as the first term in this sequence, and we derive an explicit result for the three-loop diagram (Fig.~1{\em b}) quoted in Eq.~(7) of the MT.  Finally, we will discuss the crossing diagrams in Fig.~1{\em d} and {\em e} of the MT.  
\begin{figure}
\centering
\includegraphics[scale=0.5]{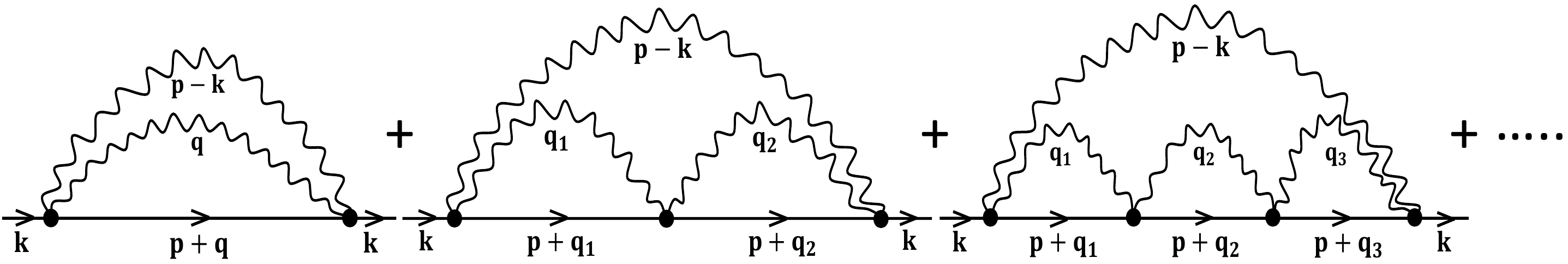}
\caption{\label{umbrella} Umbrella diagrams}
\end{figure}
\subsection{Non-crossing ``umbrella'' diagrams}
\subsubsection{Resummation of ``umbrella'' diagrams}
The sequence of umbrella diagrams is reproduced in Fig.~\ref{umbrella}. Note that the electron and phonon momenta are labeled differently compared to the MT.  As in Sec.~\ref{sec:2ph}, we treat phonons as classically occupied modes with linear dispersion. We define an effective vertex
\beq
\label{newv}
\Upsilon
_{\alpha\beta}(\bq) \equiv T\Gamma_{\alpha\beta}(\bq) D_0(q,0) = - \frac{4\pi a}{q^2} \left( \delta_{\alpha\beta} - \frac{q_\alpha q_\beta}{q^2} \right),
\eeq
where
\beq
a = \frac{T g_2 \Omega_0^2}{8\pi^2 s^2},
\eeq
$D_0$ and $\Gamma_{\alpha\beta}$ are given Eqs.~(\ref{phmat}) and (\ref{vertex}), respectively. Next, we define an effective single-phonon ``self-energy" as
\beq
\hat {\mathcal S}
(\bp, 
\e) = -\int_\bq 
\hat\Upsilon
(\bq) G
(\bp+\bq, 
\e),
\eeq
where  hats on ${\mathcal S}$ and $\Upsilon$ indicate that they are $3\times 3$ matrices, and $G
(\bp, \epsilon)
$ 
is the retarded 
Green's function.
Then the sum of the umbrella diagrams for the self-energy is given  by 
\beq
\label{self_um}
\Sigma
(\bk,\e) = - \frac{1}{2} \text{Tr} \int_\bp \hat{
\Upsilon}(\bp-\bk) \left[ \hat{{\mathcal S}} + \hat{{\mathcal S}}^2 + \hat{{\mathcal S}}^3 + ... \right] = - \frac{1}{2} \text{Tr} \int_\bp \hat{
\Upsilon}(\bp-\bk) \hat{{\mathcal S}} (\hat{I} - \hat{{\mathcal S}})^{-1}.
\eeq
The calculation is facilitated by switching to the real-space representation 
\beq
\label{real_S}
\hat {\mathcal S}_{\alpha\beta}(\bp, \e) = -\int d^3r \Upsilon_{\alpha\beta} (\br) G(-\br, \e)e^{i\bp\cdot\br}.\eeq
where $G(\br, \e)=\int_\bk e^{i\bk\cdot\br} G(\bk,\e)$ is the Green's function in real space. Since the electron dispersion is assumed to be isotropic, $G(\br, \e)=G(-\br, \e)=G(r,\e)$. 

The real-space representation of the effective vertex is found as
\bea
\Upsilon_{
\alpha\beta}&=& -a\int_\bq \frac{4\pi}{q^2}\left(\delta_{\alpha\beta}-\frac{q_\alpha q_\beta}{q^2}\right)e^{i\bq\cdot\br}=-a\left(\frac{\delta_{\alpha\beta}}{r}+\sigma_{\alpha\beta}\right)
\eea
where
\bea
\sigma_{\alpha\beta}=-4\pi \int_\bq\frac{q_\alpha q_\beta}{q^4}e^{i\bq\cdot\br}=4\pi \frac{\partial^2}{x_\alpha x_\beta}\int_\bq\frac{e^{i\bq\cdot\br}}{q^4}.
\eea
To calculate the last integral, we regularize it as 
\bea
4\pi \int_\bq\frac{e^{i\bq\cdot\br}}{(q^2+b)^2}=-4\pi \frac{\partial}{\partial b} \int_\bq\frac{e^{i\bq\cdot\br}}{q^2+a}=-\frac{\partial}{\partial b} \frac{e^{-\sqrt{b}r}}{r}=\frac{1}{2\sqrt{b}}e^{-\sqrt{b}r}
\eea
and take $b\to 0$ at the end. This yields
\beq
\sigma_{\alpha\beta}=\frac{\partial^2}{\partial x_\alpha \partial x_\beta} \frac{1}{2\sqrt{b}}e^{-\sqrt{b}r}=-\frac{\partial}{\partial x_\alpha}\left(\frac{x_\beta}{2r}e^{-\sqrt{b}r}\right)=-\left(\frac{\delta_{\alpha\beta}}{2r}-\frac{x_\alpha x_\beta}{2r^3}-\frac{\sqrt{b} x_\alpha x_\beta}{4r^2}\right)e^{-\sqrt{b}r}\Big\vert_{b\to 0}=-\frac{\delta_{\alpha\beta}}{2r}+\frac{x_\alpha x_\beta}{2r^3}
\eeq
Therefore,
\bea
\label{ecp}
\Upsilon_{\alpha\beta}=-\frac{a}{2}\left(\frac{\delta_{\alpha\beta}}{r}+\frac{x_\alpha x_\beta}{r^3}.
\right)
\eea

 \subsubsection{Umbrella diagrams in the quasiparticle regime}
 In the quasiparticle regime, the electron Green's function can be replaced by its free-fermion form. For on-shell electrons, i.e., for $\xi_\bk=\te=\e+\mu$, 
 \beq
\label{real_gr}
G(r, k) = -\frac{m^*}{2\pi r} e^{ikr},
\eeq
where $k=\sqrt{2m^*\te}+i0^+$  (Writing $k$ in this form, we took into account that $\te>0$ for mass-shell particles. In the rest of the calculations, $k$ can be treated as purely real.)
The self-energy in this regime also can be evaluated on the mass-shell. Denoting the mass-shell value of ${\mathcal S}$ via ${\mathcal S}(\bp,k)$, we decompose ${\mathcal S}(\bp,k)$ into the sum of isotropic and anisotropic parts:
\bea
{\mathcal S}_{\alpha\beta}(\bp,k)=-\frac{ma}{4\pi} \int d^3r \frac{e^{i\left(kr+\bp\cdot \br\right)}}{r}\left(\frac{\delta_{\alpha\beta}}{r}+\frac{x_\alpha x_\beta}{r^3}\right)=-\frac{ma}{4\pi}\left[ \delta_{\alpha\beta} {\mathcal S}^0(p,k)+{\mathcal S}^1_{\alpha\beta}(\bp,k)\right].
\eea
The isotropic part is given by
\bea
{\mathcal S}^0(p,k)=\int d^3 r\frac{e^{i k r}}{r^2} e^{i\bp\cdot\br}
=\frac{4\pi}{p}  I_0\left(\frac{k}{p}\right),
\eea
where 
\bea
\label{I_0}
I_0(y)= \int^\infty_0 \frac{dx}{x}e^{iyx}\sin x
=  \frac{1}{2}\left[ \pi \theta(1-y)+i\ln\Big\vert\frac{y+1}{y-1}\Big\vert\right].
\eea
The anisotropic part is given by\bea
{\mathcal S}^1_{\alpha\beta}(\bp,k)&=&\int d^3 r\frac{e^{i \left(k r+\bp\cdot\br\right)}}{r^4} x_\alpha x_\beta 
=\frac{4\pi}{p} 
\left[\left(\delta_{\alpha\beta}-3\frac{p_\alpha p_\beta}{p^2}\right)I_1\left(\frac{k}{p}\right)+
\frac{p_\alpha p_\beta}{p^2} I_0\left(\frac{k}{p}\right)\right],
\eea
where
\bea
\label{I_1}
I_1(y)=-\int^\infty_0  \frac{dx}{x} \frac{d}{dx} \frac{\sin x}{x} e^{iy x}=\frac{1}{2} (1 - y^2) I_0(y) + \frac{i}{2} y
\eea
Combining the two parts of $\hat {\mathcal S}$ together, we obtain 
\bea
\label{sphse}
{\mathcal S}_{\alpha\beta}(\bp,k)=-\frac{m^* a}{p} \left\{\delta_{\alpha\beta}\left[I_0\left(\frac{k}{p}\right)+I_1\left(\frac{k}{p}\right)\right]+\frac{p_\alpha p_\beta}{p^2}\left[I_0\left(\frac{k}{p}\right)-3I_1\left(\frac{k}{p}\right)\right]\right\}.
\eea

\paragraph{Second-order two-phonon process.} For completeness, we re-derive the result for the two-loop self-energy in the perturbative regime (to second order in $g_2$) by using the representation developed in the last  section. To the specified order in $g_2$, 
\beq
\Sigma(k)\equiv \Sigma(\bk,\te=\xi_\bk) = -\frac{1}{2} \text{Tr} \int_\bp \hat{\Upsilon}(\bp-\bk) \hat{{\mathcal S}}(\bp,k).
\eeq
Choosing the polar axis of a spherical coordinate system along $\bk$ and integrating over the azimuthal angle, we obtain for  the trace  in Eq.~(\ref{sphse}) 
\beq
\int_0^{2\pi} \frac{d\phi}{2\pi} \text{Tr} [...] = \frac{4\pi m^* a^2}{p |\bp - \bk|^2} \left[ 2(I_0+I_1) + \frac{k^2 \sin^2\theta}{|\bp - \bk|^2} (I_1-3I_1) \right].
\eeq
Next, the polar-angle integration yields 
\beq
\label{self_2ph}
\Sigma(k) = \frac{m^* a^2}{4\pi k} \int_0^\infty dp \left[ -4(I_0+I_1) \text{ln} \left|\frac{p+k}{p-k} \right| + \frac{2k}{p}(I_0-3I_1) \left( 1 - \frac{p^2+k^2}{2pk} \text{ln} \left| \frac{p+k}{p-k} \right| \right) \right].
\eeq 
The imaginary part of Eq.~\ref{self_2ph} is
read off as
\beq
\Sigma'' = \frac{m^*a^2}{4\pi} {\cal C}_2
\eeq
where $x 
=k/p
$ and
\beq
{\cal C}_2= \int_0^\infty dx \left[ -4\left( \frac{1}{2x} + \frac{3x^2-1}{4x^2} \text{ln} \left| \frac{x+1}{x-1} \right| \right) \text{ln} \left| \frac{x+1}{x-1} \right| + \frac{2}{x} \left( -\frac{3}{2x} + \frac{3-x^2}{4x^2} \text{ln} \left| \frac{x+1}{x-1} \right| \right) \left( 1 - \frac{x^2+1}{2x} \text{ln}\left| \frac{x+1}{x-1} \right| \right) \right]
=3\pi.
\eeq
 Defining the (quantum) relaxation time as  $1/\tau_0 \equiv -2\Sigma''$, we get
\beq
\frac{1}{\tau_0} = \frac{3T^2}{128 \pi^3} \frac{m^* g_2^2 \Omega_0^4}{s^4} = \frac{3T^2}{2E_0}.
\eeq
which coincides with Eq.~(\ref{sctt}), obtained in a different way.

\paragraph{Third-order two-phonon process.}
In this section, we will calculate the correction to the scattering rate due to a third-order two-phonon process, depicted  by diagram  {\em b} Fig.~$1$ of the MT  or, equivalently, by the middle diagram in Fig.~\ref{umbrella}. The  third-order self-energy reads:
\beq
\Sigma(k) = -\frac{1}{2} \text{Tr} \int_\bp \hat{\Upsilon}(\bp-\bk) \hat{{\mathcal S}}^2(\bp,k).
\eeq
Integrating the trace in the equation above over the azimuthal angle, we find
\beq
\int_0^{2\pi} \frac{d\phi}{2\pi} \text{Tr}[...] = -\frac{4\pi {m^*}^2 a^3}{p^2 |\bp-\bk|^2} \left[ 2(I_0+I_1)^2 + \frac{k^2 \sin^2\theta}{|\bp-\bk|^2} (I_0-3I_1)(3I_0-I_1)  \right].
\eeq
The integral over the polar angle yields \beq
\label{self_3ph_2}
\Sigma(k) = -\frac{{m^*}^2 a^3}{2\pi k} \int_0^\infty \frac{dp}{p} \left[ -2 (I_0+I_1)^2 \text{ln}\left| \frac{p+k}{p-k} \right| + \frac{k}{p} (I_0-3I_1)(3I_0-I_1) \left( 1 - \frac{p^2+k^2}{2pk} \text{ln} \left| \frac{p+k}{p-k} \right| \right) \right].
\eeq
The imaginary part of Eq.~(\ref{self_3ph_2}) is read off as
\beq
\begin{split}
\Sigma''(k) &= \frac{{m^*}^2 a^3}{2k} \int_0^\infty \frac{dx}{x} \Theta(x-1) \Bigg[ \frac{3x^2-1}{2x^2} \left( \frac{1}{x} + \frac{3x^2-1}{2x^2} \text{ln} \left| \frac{x+1}{x-1} \right| \right) \text{ln} \left| \frac{x+1}{x-1} \right| \\
&\hspace{3cm} + \frac{1}{x} \left( \frac{7x^2+3}{4x^3} + \frac{(5x^2+1)(x^2-3)}{8x^4} \text{ln} \left| \frac{x+1}{x-1} \right| \right) \left( 1 - \frac{x^2+1}{2x} \text{ln} \left| \frac{x+1}{x-1} \right| \right) \Bigg] \\
&= \frac{{m^*}^2 a^3}{64k} [6+217\zeta(3)] = \frac{[6+217\zeta(3)]}{64\pi^{3/2}} \frac{T^3 \sqrt{m^*}}{E_0^{3/2} k}.
\end{split}
\eeq
The third-order correction to the scattering rate is thus given by
\beq
\label{tau3}
\delta\left(\frac{1}{\tau}\right) = -\frac{[6+217\zeta(3)]}{32\pi^{3/2}} \frac{T^3 \sqrt{m^*}}{E_0^{3/2} k} \approx -1.50 \frac{T^3 \sqrt{m^*}}{E_0^{3/2} k}.
\eeq
Note that the correction is negative, which is also the case for higher-order processes. 
This means that multi-phonon processes reduce the scattering rate, which is consistent with the result that $\rho$ in the non-quasiparticle regime varies with $T$ slower than  in the quasiparticle one: as $T^{3/2}$ vs $T^2$. 

The calculation of the transport scattering rate due at third order is too lengthy to be presented here. The result is that the transport scattering rate is the same as in Eq.~(\ref{tau3}), except for  the numerical coefficent of 1.50 is replaced by 1.24. This is the result quoted in Eq.~(7) of the MT.

\subsubsection{Non-quasiparticle regime}
In the previous section we showed that higher-order umbrella diagrams are parametrically smaller than the two-phonon one in the quasiparticle regime; for non-degenerate electrons, the small parameter is $\sqrt{T/E_0}$. This indicates that the contribution of high-order diagrams becomes important at $T\sim E_0$ and, in principle, one has to solve Eq.~(\ref{self_um}) in the non-quasiparticle regime self-consistently. However, this presents a daunting task, both analytically and numerically. In the MT and in Secs.~\ref{sec:nqp} and \ref{sec:num}, we solved the self-consistent equation that corresponds to the first, two-phonon diagram diagram in Fig.~\ref{umbrella} or, equivalently, to the first term in the expansion of Eq.~(\ref{self_um}) back in  $\hat {\mathcal S}$. In this section, we will argue that higher-order terms in $\hat {\mathcal S}$ do change the result by $\mathcal {O}(1)$ but not more than that.  Since numerical coefficients do not play any role in the forthcoming analysis, we adopt a simplified ``scalar'' model, in which  the effective tensor vertex in Eq.~(\ref{gamma}) is replaced by its diagonal part
\bea
\Upsilon_{\alpha\beta}(\bq)\to \Upsilon_0(\bq)\delta_{\alpha\beta}.
\eea
where $\Upsilon_0(\bq)= -\frac{4\pi a}{q^2}$.
It can be checked that the scalar model reproduces all the results of the previous sections except for a change in the numerical coefficients. Accordingly, Eq.~(\ref{self_um}) is replaced by
\beq
\label{se_um}
\Sigma(\bk, \epsilon) = -\int_\bp \Upsilon_0(\bp - \bk)  \frac{{\mathcal S}(\bp, \epsilon)}{1-{\mathcal S}(\bp, \epsilon)},
\eeq
where
\beq
{\mathcal S}(\bp, \epsilon) = -\int_\bq G(\bp+\bq, \epsilon) \Upsilon_0(\bq)
\eeq
is now a scalar.
Relabeling $\bp+\bq=\bk'$ and taking the imaginary part of Eq.~(\ref{se_um}), we obtain
\beq
\label{se_um1}
\Sigma''(\bk, \epsilon) = \int_{\bk'\bq} \Upsilon_0(\bk'-\bk-\bq) \Upsilon_0(\bq) \frac{\Sigma''(\bk')}{\left[ \te - \xi_{\bk'} - \Sigma'(\bk', \epsilon) \right]^2 + \left[ \Sigma''(\bk', \epsilon) \right]^2} \underbrace{\left[ \frac{1}{\left[ 1 - {\mathcal S}'(\bk'-\bq, \epsilon) \right]^2 + \left[ {\mathcal S}''(\bk'-\bq, \epsilon) \right]^2} \right]}_{\equiv R}.
\eeq
If the last factor in square brackets on the RHS of Eq.~(\ref{se_um1}), symbolically denoted by $R$,  is replaced by 1, we are back to two-loop order. In the previous sections, we showed the two-loop self-energy exhibits a threshold behavior as a function of $\e$ and that transport in the non-quasiparticle regime is controlled by energies near the threshold. In terms of variable $\bar\e=-\e-\mu$, the threshold is at $\e_0\sim \lambda\sim T^2/E_0$.  Near the threshold, 
$\Sigma''$ is small and the threshold position can be determined from the self-consistent equation linearized in $\Sigma''$. Since ${\mathcal S}''$ and $\Sigma''$ are proportional to each other, the $R$ factor does not contribute extra $\mathcal{O}(\Sigma'')$  terms. At the two-loop level, we also assumed that $\Sigma'$ is a slowly varying function of its arguments and absorbed it into $\e$. If we apply the same approximation to ${\mathcal S}'$, the linearized equation for $\Sigma''$ is reduced back to the two-loop form:
\beq
\label{se_um3}
\Sigma''(\bk, \bar{\epsilon}) = \int_{\bk', \bq} \Upsilon_0(\bk' - \bk - \bq) \Upsilon_0(\bq) \frac{\Sigma''(\bk', \bar{\epsilon})}{(\bar{\epsilon} + \xi_{\bk'})^2}.
\eeq
Therefore, in this approximation higher-order diagrams do not affect the threshold position.

The behavior near the threshold is obtained by expanding the non-linear equation (\ref{se_um1}) to ${\mathcal O}\left([\Sigma'']^3\right)$. To two-loop order, the result for $\gamma_{\e}(\xi)=-\Sigma''(\xi,\e)$ is given by Eq.~(\ref{gamma}). Although in this case the scaling function $S(x)$ is known [cf. Eq.~(\ref{S})], its explicit form determines only numerical coefficient $C$ in Eq.~(\ref{res_largeT}) for the resistivity, whereas the $T^{3/2}$ scaling form of $\rho$ is determined by the square-root threshold singularity in $\gamma_{\e}(\xi)$ as a function of $\e_0-\bar\e$. We will now argue that the threshold singularity is not affected by higher loop terms. Indeed, by expanding the $R$ factor in Eq.~(\ref{se_um1}) to order $\mathcal{O}\left([\Sigma'']^2\right)$, we add another $\mathcal{O}\left([\Sigma'']^3\right)$ term to the RHS of Eq.~(\ref{thresh1}). Because the linear-order result is still given by Eq.~(\ref{gamma1}), the linear proportionality between $C^2(\bar\e)$ and $\delta\e=\bar\e-\e_0$ still remains in place even in the presence of an extra $\mathcal{O}\left([\Sigma'']^3\right)$ term. This implies that $\gamma_{\e}(\xi)=\sqrt{(\bar\e_0-\bar\e)\e_0}\times\tilde S(\xi/\e_0)$ with some new scaling function $\tilde S(x)$ which, as we said above, determines only a numerical coefficient in the resistivity.

In reality, both $\Sigma'$ and ${\mathcal S}'$ vary with $\e$ and $\xi$. Because $\Sigma'$ varies logarithmically with its arguments, one expects the linear dependence of threshold $\e_0$ on coupling constant $\lambda$ to change to $\lambda\ln\lambda$, which is what we see in the numerical calculation, cf. Fig.~\ref{pos_thr}. To estimate the effect of ${\mathcal S}'$ being a function of its arguments, we retain ${\mathcal S}'$ in the linearized equation, which then becomes
\beq
\label{se_um2}
\Sigma''(\bk, \bar{\epsilon}) = \int_{\bk', \bq} \Upsilon_0(\bk' - \bk - \bq) \Upsilon_0(\bq) \frac{\Sigma''(\bk', \bar{\epsilon})}{(\bar{\epsilon} + \xi_{\bk'})^2} \left[ \frac{1}{\left[ 1 - {\mathcal S}'(\bk'-\bq) \right]^2} \right],
\eeq
where
\beq
{\mathcal S}'(\bp, \bar{\epsilon}) = \int_{\bp'} \Upsilon_0(\bp-\bp') \frac{1}{\bar{\epsilon} + \xi_{\bp'}}\sim \frac{T}{\sqrt{E_0\xi}} f\left(\frac{\bar\e}{\xi}\right),
\eeq
where $\xi\equiv \xi_\bp$ and
\beq
\label{S}
f(x)=\int_0^\infty dy \frac{1}{y + x} \text{ln}\left| \frac{\sqrt{y} + 1}{\sqrt{y} - 1} \right|,
\eeq
such that $f(x\to 0)=\pi^2$ and $f(x\to\infty)=2\pi/\sqrt{x}$. In the non-quasiparticle regime, transport is controlled by the region of $\bar\e\approx \e_0\sim T^2/E_0$ and $\xi\sim \e_0\sim T^2/E_0$. In this region, ${\mathcal S}'\sim 1$, and therefore the threshold position can be shifted by ${\mathcal O}(1)$ but not more than that.

\subsection{Crossing diagrams}
\begin{figure}
\centering
\includegraphics[scale=0.5]{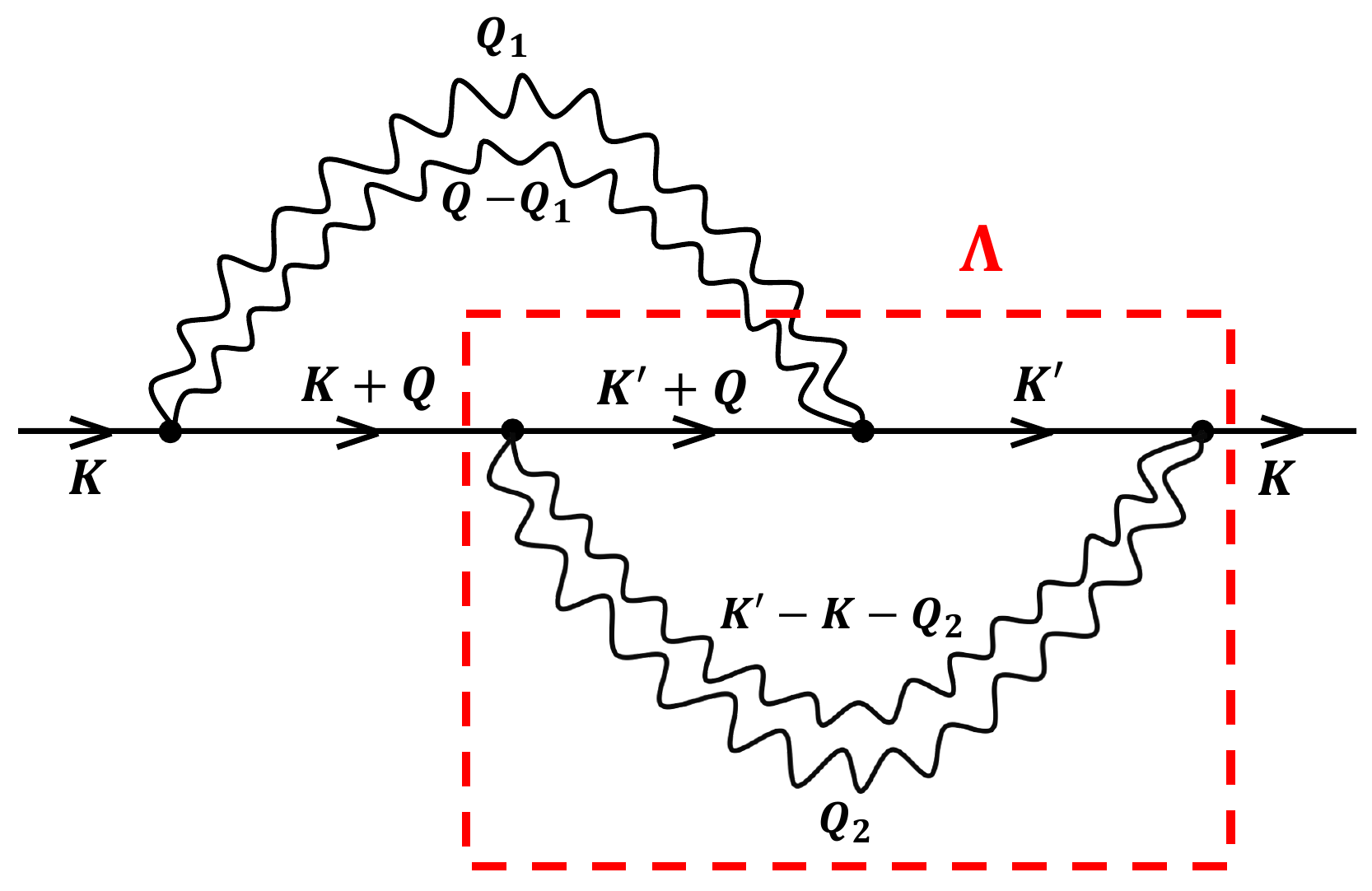}
\caption{\label{fig:cross} Fourth-order crossing diagram.}
\end{figure}

In this section, we estimate the contribution from the crossing fourth-order two-phonon diagram, Fig.~1{\em e} of the MT, which is reproduced in Fig.~\ref{fig:cross} for reader's convenience. This diagram is a vertex correction to the second-order process, Fig.~1{\em a} of the MT or, equivalently, the first diagram in Fig.~\ref{umbrella} of this document.  An analogous vertex correction for single-phonon scattering is small by the Migdal theorem, \cite{migdal} which applies in the adiabatic regime. In what follows, we generalize Migdal's argument for two-phonon scattering. 

 We are interested in the magnitude of  triangular vertex $\Lambda$, as defined in Fig.~\ref{fig:cross}. This vertex is dimensionless and hence needs to be compared to 1. Also, because the vertex involves virtual states, it can be calculated at $T=0$. Finally, as we are interested only in an order-of-magnitude estimate, we again adopt a scalar model, in which the tensor vertex in Eq.~(\ref{vertex}) is replaced by its diagonal part
\bea
\Gamma_{\alpha\beta}(\bq)\to \Gamma_0(\bq)\delta_{\alpha\beta},
\eea
where $\Gamma_0(\bq)=g_2\Omega_0^2/4\pi sq$ for massless phonons.
After these steps, $\Lambda$ is written as
\beq
\label{tri}
\Lambda = \int d^4 K' \int d^4 Q_2 \Gamma_0 (\bq_2) \Gamma_0(\bk' - \bk - \bq_2) D_0(Q_2) D_0(K'-K-Q_2) G_0(K' + Q) G_0(K'),
\eeq
where $P=(\bp,i\omega_m)$, $d^4P = \int d\omega_m\int d^3p/(2\pi)^4$, $G_0(P)=(i\omega_m-\xi_\bp+\mu)$ and $D_0(P)$ is given by Eq.~(\ref{phmat}). It can be checked that all the momenta are on the order of incoming electron momentum $k$, whereas phonon frequencies are on the order of $sk$. For non-degenerate electrons, $k\approx k_F$ and $k\gg m^*s$ by the adiabatic condition, $v_F\gg s$, which we assume to be satisfied. For degenerate electrons, $k\sim \sqrt{m^*T}$ and $k\gg m^*s$ for $T\gg m^*s^2\sim 5$\,K in STO, which we also assume to be satisfied. Therefore, $k\gg m^*s$  for all cases of interest, which implies that the typical phonon frequencies, $sk$, are small compared to electron energy $\xi_\bk=k^2/2m$. Neglecting phonon frequencies in the electron Green's functions, we estimate the latter as $G_0 \sim 1/\xi_\bk$, while $D_0 \sim 1/sk$. The integrals over momenta in Eq.~(\ref{tri}) give a factor of $(k^3)^2$ while those over frequencies give a factor of $(sk)^2$. Then $\Lambda$ is  estimated as
\beq
\Lambda \sim g_2^2 \Omega^4 (s^2 k^2) (k^6) \frac{1}{\xi_\bk^2} \frac{1}{s^2 k^2} \frac{1}{s^2 k^2}\sim \frac{m^*s^2}{E_0},
\eeq
where $E_0$ is defined in Eq.~(\ref{E0}). We see that the effective Migdal parameter is small if $m^*s^2\ll E_0$, which is satisfied with a large margin because $m^*s^2\sim 5$\,K while $E_0\sim 200$\,K.
Therefore, the crossing diagram gives only a small correction to the second-order result. 
\section{Boltzmann equation for two-phonon scattering}
\label{app:T2T5}
In the MT, we found the resistivity due to two-TO-phonon scattering in the classical regime, when phonons are treated as static disorder. In this regime, one can simply calculate the corresponding transport time and substitute it into the Drude formula.  In this section, we consider the Boltzmann equation which allows one to analyze both the classical and Bloch-Gr{\"u}neisen regimes. The latter is  defined by the condition $\omega_0\ll T\ll T_{\text{BG}}=2k_Fs$. The adiabaticity condition, $s\ll v_F$, implies that   $T_{\text{BG}}\ll E_F$, i.e., that electron statistics is degenerate. On the other hand, the 
 condition
 $T\gg \omega_0$ 
 allows one to replace the TO phonon spectrum, given by Eq.~\ref{TOsp},
 by an acoustic one: $\omega_\bq=sq$. 
 We solve the Boltzmann equation here under one additional assumption, i.e., we assume that $T$ is low enough such that impurities provide the dominant scattering mechanism. In this case, the electron-phonon collision integral can be treated as a correction to the electron-impurity one.   In a low carrier-density system,  the Bloch-Gr{\"u}neisen regime sets in at low temperatures, and thus the assumption of dominant impurity scattering is reasonable.

The collision integral for two-phonon scattering  is given by
\beq
\label{coll-2ph}
\begin{split}
I_{\text{coll}}^{2\text{ph}} = \sum_{\bp', \bq_1, \bq_2} \bigg[ & W(\bp', \bq_1, \bq_2; \bp) \Big( n_{\bp'} (1 - n_{\bp}) N_{\bq_1} N_{\bq_2} - n_{\bp} (1 - n_{\bp'}) (1 + N_{\bq_1}) (1 + N_{\bq_2}) \Big) \delta(\xi_{\bp} - \xi_{\bp'} - \omega_{\bq_1} - \omega_{\bq_2}) \\
+& W(\bp', \bq_1; \bp, \bq_2) \Big( n_{\bp'} (1 - n_{\bp}) N_{\bq_1} (1 + N_{\bq_2}) - n_{\bp} (1 - n_{\bp'}) (1 + N_{\bq_1}) N_{\bq_2} \Big) \delta(\xi_{\bp} - \xi_{\bp'} - \omega_{\bq_1} + \omega_{\bq_2}) \\
+& W(\bp', \bq_2; \bp, \bq_1) \Big( n_{\bp'} (1 - n_{\bp}) (1 + N_{\bq_1}) N_{\bq_2} - n_{\bp} (1 - n_{\bp'}) N_{\bq_1} (1 + N_{\bq_2}) \Big) \delta(\xi_{\bp} - \xi_{\bp'} + \omega_{\bq_1} - \omega_{\bq_2}) \\
+& W(\bp'; \bp, \bq_1, \bq_2) \Big( n_{\bp'} (1 - n_{\bp}) (1 + N_{\bq_1}) (1 + N_{\bq_2}) - n_{\bp} (1 - n_{\bp'}) N_{\bq_1} N_{\bq_2} \Big) \delta(\xi_{\bp} - \xi_{\bp'} + \omega_{\bq_1} + \omega_{\bq_2}) \bigg].
\end{split}
\eeq
Here $W(S_1;S_2)$ denotes the scattering probability between states $S_1$ and $S_2$ specified by the electron and phonon momenta, and
$n_{\bp} = n_{F}(\xi_\bp) + \delta n_{\bp}$ and $N_{\bq} = n_{B}(\omega_\bq) + \delta N_{\bq}$ are the distribution functions of electrons and phonons, respectively, with $n_{F}$ and $n_{B}$ being the Fermi and Bose equilibrium distributions.
It is understood that the scattering probability includes the constraint imposed by momentum conservation, e.g.,
\bea
 W(\bp', \bq_1, \bq_2; \bp)=\tilde W\delta(\bp'-\bp-\bq_1-\bq_2),
\eea
etc.
 To lowest order in electron-phonon interaction,  kernel $\tilde W$  is same in all the scattering probabilities and related to the vertex in Eq.~(\ref{vertex}) via
 \bea
 \tilde W=2\times 2\pi \sum_{\alpha\beta}\frac{1}{4}\Gamma_{\alpha\beta}(\bq_1)\Gamma_{\beta\alpha}(\bq_2)=\frac{
 g_2^2}{
 16
 \pi
 } \frac{
 \Omega_0^4}{\omega_{\bq_1} \omega_{\bq_2}} \bigg( 1 + \frac{(\bq_1 \cdot \bq_2)^2}{q_1^2 q_2^2} \bigg).
 \eea
A factor of 2 in the first part of the equation above is the combinatorial coefficient of the two-phonon diagrammatic technique, as explained in Sec.~\ref{sec:2ph}, and a factor of $1/4$ is related to our definition of the coupling constant in Eq.~(1) of the MT.
 
The phonons are assumed to be in the equilibrium state, such that  $\delta N_{\bq}=0$.
The non-equilibrium electron distribution is parameterized as $\delta n_\bp = - \phi_\bp \partial n_{F}/\partial \xi_\bp$.
After linearizing in $\phi_\bk$, the Boltzmann equation is reduced to
\bea
\frac{e(\bp \cdot \bE)}{m^*} \frac{\partial n_F}{\partial \xi} &=& \frac{\partial n_F}{\partial \xi} \frac{\phi}{\tau_i}+\frac 1T\int \frac{d^3 q_1}{(2\pi)^3} \int \frac{d^3 q_2}{(2\pi)^3} 
\tilde W
A(\bp,\bq_1,\bq_2)
\eea
where  the first term on the right-hand side (RHS) is the electron-impurity collision integral with scattering time $\tau_i$, $\tilde W$ is the scattering probability after momentum-conservation constraint is resolved, and
\bea
A(\bp,\bq_1,\bq_2)&&=
\Delta\phi_{\bp-\bq_1-\bq_2,\bp} n_B(\omega_1 + \omega_2) \left[1 + n_{B}(\omega_1)\right] \left[1 + n_{B}(\omega_2)\right] 
 \left[n_F(\xi_{\bp}-\omega_1-\omega_2)-n_F(\xi_\bp)\right]
 \delta(\xi_\bp - \xi_{\bp-\bq_1-\bq_2}  - \omega_1 - \omega_2)\nn\\
&&+\Delta\phi_{\bp-\bq_1+\bq_2,\bp}n_B(\omega_1 - \omega_2) \left[1 + n_{B}(\omega_1)\right]n_{B}(\omega_2)
 \left[n_F(\xi_{\bp}-\omega_1+\omega_2)-n_F(\xi_\bp)\right]
\delta(\xi_\bp - \xi_{\bp-\bq_1+\bq_2} - \omega_1 + \omega_2)\nn\\ 
&&+\Delta\phi_{\bp+\bq_1-\bq_2,\bp} n_B(- \omega_1 + \omega_2) n_{B} \left[1 + n_{B}(\omega_2)\right]
 \left[n_F(\xi_{\bp}+\omega_1-\omega_2)-n_F(\xi_\bp)\right]
\delta(\xi_\bp - \xi_{\bp+\bq_1-\bq_2}  + \omega_1 - \omega_2)\nn \\
&&+ \Delta\phi_{\bp+\bq_1+\bq_2,\bp}n_B(- \omega_1 - \omega_2) n_{B}(\omega_1) n_{B}(\omega_2)
 \left[n_F(\xi_{\bp}+\omega_1+\omega_2)-n_F(\xi_\bp)\right]
\delta(\xi_\bp - \xi_{\bp+\bq_1+\bq_2}+ \omega_1 + \omega_2),
\label{appbe1}
\eea
where $\Delta\phi_{\bk,\bk'}=\phi_{\bk}-\phi_{\bk'}$ and  
$\omega_i\equiv \omega_{\bq_i}$ ($i=1,2$).
We write $\phi_\bk = \phi_\bk^0+ \phi_\bk^1$, where $\phi_0=e(\bk \cdot \bE) \tau_i/Tm^*$  is the solution of (\ref{appbe1}) without the electron-phonon collision integral, and  $ \phi_\bk^1$ is a correction found from 
\bea
\label{be}
-\frac{\partial n_F}{\partial \xi} \phi^1_\bp &=& \frac{e \tau_i^2}{m^*T} \int \frac{d^3 q_1}{(2\pi)^3} \int \frac{d^3 q_2}{(2\pi)^3}
W
A^0(\bp,\bq_1,\bq_2),
\eea
where $A^0(\bp,\bq_1,\bq_2)$ is obtained from $A^0(\bp,\bq_1,\bq_2)$ in Eq.~\eqref{appbe1} by replacing all $\phi$'s by $\phi^0$.
A correction to the electric current due to electron-phonon interaction is found as
\beq
\delta \bj = 2e\int \frac{d^3 p}{(2\pi)^3} \bigg( -\frac{\partial n_F}{\partial \xi} \bigg) \phi_\bp^1 \frac{\bp}{m^*} = eN_F \int d\xi_{\bp} \int \frac{d \Omega_{\bp}}{4\pi} \bigg( -\frac{\partial n_F}{\partial \xi} \bigg) \phi_\bp^1 \frac{\bp}{m^*},
\eeq
where $N_F$ is the density of states at the Fermi energy. To necessary accuracy,  $\bp$ can be replaced $\bp_F=k_F\bp/p$ 
anywhere but in the Fermi functions. Using an identity $\int d\xi \left[n_F(\xi-\epsilon)-n_F(\xi)\right]=\epsilon$, we perform the integrals over $\xi_\bp$ and arrive at the following expression for the correction to the conductivity
\beq
\label{be2}
\begin{split}
\delta \sigma = - \frac{e^2 N_F \tau_i^2}{Tm^{*2} E^2} \int \frac{\Omega_{\bp}}{4\pi} & \int \frac{d^3 q_1}{(2\pi)^3} \int \frac{d^3 q_2}{(2\pi)^3} \tilde W(\bp_F \cdot \bE) \times \\
& \times \Big\{ (\omega_1 + \omega_2) (\bq_1 + \bq_2) \cdot \bE \Big( n_B(\omega_1 + \omega_2) \left[1 + n_{B}(\omega_1)\right] \left[1 + n_{B}(\omega_2)\right] \delta(\xi_{\bp_F} - \xi_{\bp_F - \bq_1 - \bq_2} - \omega_1 - \omega_2) \\
&\hspace{3cm} - \left[1 + n_B(\omega_1 + \omega_2)\right] n_{B}(\omega_1) n_{B}(\omega_2) \delta(\xi_{\bp} - \xi_{\bp_F + \bq_1 + \bq_2} + \omega_1 + \omega_2) \Big) \\
&\hspace{0.2cm}+ (\omega_1 - \omega_2) (\bq_1 - \bq_2) \cdot \bE \Big( n_B(\omega_1 - \omega_2) \left[1 + n_{B}(\omega_1)\right] n_{B}(\omega_2) \delta(\xi_{\bp_F} - \xi_{\bp_F - \bq_1 + \bq_2} - \omega_1 + \omega_2) \\
&\hspace{3cm} - \left[1 + n_B(\omega_1 - \omega_2)\right] n_{B}(\omega_1) \left[1 + n_{B}(\omega_2)\right] \delta(\xi_{\bp_F} - \xi_{\bp_F + \bq_1 - \bq_2} + \omega_1 - \omega_2) \Big) \Big\}.
\end{split}
\eeq
Upon changing $\bq_1 \rigt -\bq_1$, the last two lines in Eq.~\eqref{be2}, corresponding to simultaneous absorption and emission, are written as
\beq
\begin{split}
-(\omega_1 - \omega_2) (\bq_1 + \bq_2) \cdot \bE \Big( & n_B(\omega_1 - \omega_2) \left[1 + n_{B}(\omega_1)\right] n_{B}(\omega_2) \delta(\xi_{\bp_F} - \xi_{\bp_F + \bq_1 + \bq_2} - \omega_1 + \omega_2) \\
-& \left[1 + n_B(\omega_1 - \omega_2)\right] n_{B}(\omega_1) \left[1 + n_{B}(\omega_2)\right] \delta(\xi_{\bp_F} - \xi_{\bp_F - \bq_1 - \bq_2} + \omega_1 - \omega_2) \Big).
\end{split}
\eeq
Changing to a new set of variables  $(\bq, \bq_1)$, where $\bq = \bq_1 + \bq_2$,  we re-write the  delta-functions as
\beq
\label{delta}
\begin{split}
\delta(\xi_{\bp_F} - \xi_{\bp_F - \bq} - \omega_{\bq_1} - \omega_{\bq - \bq_1}) &= \frac{1}{v_F q} \delta \bigg( \cos\theta_{\bp\bq} - \frac{q}{2k_F} - \frac{s}{v_F} \frac{q_1}{q} - \frac{s}{v_F} \frac{|\bq - \bq_1|}{q} \bigg) \\
\delta(\xi_{\bp_F} - \xi_{\bp_F + \bq} + \omega_{\bq_1} + \omega_{\bq - \bq_1}) &= \frac{1}{v_F q} \delta \bigg( \cos\theta_{\bp\bq} + \frac{q}{2k_F} - \frac{s}{v_F} \frac{q_1}{q} - \frac{s}{v_F} \frac{|\bq - \bq_1|}{q} \bigg) \\
\delta(\xi_{\bp_F} - \xi_{\bp_F + \bq} - \omega_{\bq_1} + \omega_{\bq - \bq_1}) &= \frac{1}{v_F q} \delta \bigg( \cos\theta_{\bp\bq} + \frac{q}{2k_F} + \frac{s}{v_F} \frac{q_1}{q} - \frac{s}{v_F} \frac{|\bq - \bq_1|}{q} \bigg) \\
\delta(\xi_{\bp_F} - \xi_{\bp_F - \bq} + \omega_{\bq_1} - \omega_{\bq - \bq_1}) &= \frac{1}{v_F q} \delta \bigg( \cos\theta_{\bp\bq} - \frac{q}{2k_F} + \frac{s}{v_F} \frac{q_1}{q} - \frac{s}{v_F} \frac{|\bq - \bq_1|}{q} \bigg).
\end{split}
\eeq

Now we have two regimes to consider. In the first regime, the two last terms under any of the delta-functions in the equation above can be neglecting compared to $q/2k_F$. Making an assumption (to be justified {\em a posteriori}) that typical $q$ and $q_1$ in the regime are given by $\sim\max\{k_F,T/s\}$, we see that this regime corresponds to the condition $T\gg ms^2$.
Next, we use that $\cos\theta_{\bq\bE} = \cos\theta_{\bp\bE} \cos\theta_{\bp\bq} + \sin\theta_{\bp\bE} \sin\theta_{\bp\bq} \cos(\varphi_{\bp} - \varphi_{\bq})$, where $\theta_{\bk\bE}$ is a polar angle of $\bk$, measured with respect to $\bE$, and  $\varphi_{\bk}$ is the azimuthal angle of $\bk$.  The second term in $\cos\theta_{\bq\bE}$ vanishes upon the integration over the azimuthal angle, whereas a factor of $\cos\theta_{\bp\bq}$ is replaced by $\pm q/2k_F$,  depending on whether  absorption or emission processes are considered. Along with an extra factor of $q$ in Eq.~\eqref{be2}, this gives a transport factor proportional to $q^2$.

Eliminating the delta-functions and writing $\bq_2 = \bq - \bq_1$, we obtain
\beq
\label{be3}
\begin{split}
&\delta \sigma = - \frac{
g_2^2 e^2 N_F 
 \tau_i^2 
 \Omega_0^4}{
1536\pi^5  m^{*2} v_F T
} \int_0^{2k_F} dq q^3 \int_0^\infty dq_1 q_1^2 \int_{-1}^1 d(\cos\theta_{\bq\bq_1}) \frac{1}{\omega_{\bq_1}} \frac{1}{\omega_{\bq - \bq_1}} \bigg( 1 + \frac{(\bq \cdot \bq_1 - q_1^2)^2}{q_1^2 |\bq - \bq_1|^2} \bigg) \times \\
&\hspace{0.3cm} \times \bigg\{ 
(\omega_{\bq_1} + \omega_{\bq - \bq_1})
\Big[ n_B(\omega_{\bq_1} + \omega_{\bq - \bq_1})\left[1 + n_{B}(\omega_{\bq_1})\right] \left[1 + n_{B}(\omega_ {\bq - \bq_1})\right] + \left[1 + n_B(\omega_{\bq_1} + \omega_{\bq - \bq_1})\right] n_{B}(\omega_ {\bq_1}) n_{B}(\omega_{\bq - \bq_1}) \Big] \\
&\hspace{0.4cm} + 
(\omega_{\bq_1} - \omega_{\bq - \bq_1}
) \Big[ n_B(\omega_{\bq_1} - \omega_{\bq - \bq_1})\left[1 + n_{B}(\omega_{\bq_1})\right] n_{B}(\omega_ {\bq - \bq_1}) + \left[1 + n_B(\omega_{\bq_1} - \omega_{\bq - \bq_1})\right ]n_{B}(\omega_{\bq_1})\left[1 + n_{B}(\omega_{\bq - \bq_1})\right] \Big] \bigg\},
\end{split}
\eeq
where $\theta_{\bq\bq_1}$ is the angle between $\bq$ and $\bq_1$.
In a dimensionless form,  Eq.~\eqref{be3} is re-written as
\beq
\label{be4}
\begin{split}
\delta \sigma =& - T^5 \frac{g_2^2 e^2 N_F 
 \tau_i^2 
\Omega_0^4}{
1536\pi^5 m^{*2} s^7 v_F 
} \int_0^{T_{\text{BG}}/T} dx x^3 \int_0^\infty dy y \int_{-1}^1 d(\cos\theta_{\bx\by}) \bigg( 1 + \frac{(\bx \cdot \by - y^2)^2}{y^2 |\bx - \by|^2} \bigg) \times \\
&\hspace{0.3cm} \times \bigg[ \frac{y + |\bx - \by|}{|\bx - \by|} \Big( n_B(y + |\bx - \by|)(1 + n_B(y)) (1 + n_B(|\bx - \by|)) + (1 + n_B(y + |\bx - \by|)) n_B(y) n_B(|\bx - \by|) \Big) \\
&\hspace{0.5cm} + \frac{y - |\bx - \by|}{|\bx - \by|} \Big( n_B(y - |\bx - \by|)(1 + n_B(y)) n_B(|\bx - \by|) + (1 + n_B(y - |\bx - \by|)) n_B(y) (1 + n_B(|\bx - \by|)) \Big) \bigg],
\end{split}
\eeq
where 
$\bx = \bq s/T$ and $\by = \bq_1 s/T$.  Using $\delta \rho = - \rho_0^2 \delta \sigma$, where $\rho_0 = m/ne^2 \tau_i$ is the residual resistivity, and defining the effective transport rate due to electron-phonon scattering via $\delta\rho=m^*/n e^2\tau_{\text{tr}}$, we find 
\beq
\label{be5}
\begin{split}
\frac{1}{\tau_{
\text{tr}}} =& T^5 \frac{m^* g_2^2 
\Omega_0^4}{2s^4 (T_{\text{BG}})^3 (2\pi)^5} \int_0^{T_{\text{BG}}/T} dx  x ^3F(x)
\end{split}
\eeq
where
\beq
\begin{split}
F(x)=& \int_0^\infty dy y \int_{-1}^1 d(\cos\theta_{\bx\by}) \bigg( 1 + \frac{(\bx \cdot \by - y^2)^2}{y^2 |\bx - \by|^2} \bigg) \times \\
&\hspace{0.3cm} \times \bigg[ \frac{y + |\bx - \by|}{|\bx - \by|} \Big( n_B(y + |\bx - \by|)(1 + n_B(y)) (1 + n_B(|\bx - \by|)) + (1 + n_B(y + |\bx - \by|)) n_B(y) n_B(|\bx - \by|) \Big) \\
&\hspace{0.5cm} + \frac{y - |\bx - \by|}{|\bx - \by|} \Big( n_B(y - |\bx - \by|)(1 + n_B(y)) n_B(|\bx - \by|) + (1 + n_B(y - |\bx - \by|)) n_B(y) (1 + n_B(|\bx - \by|)) \Big) \bigg].
\end{split}
\eeq
Equation~\eqref{be5} is valid for any ratio of $T/T_{\text{BG}}$ but only for $T\gg m^*s^2$.

For $T \ll T_{\text{BG}}$, the upper limit of $x$-integration in Eq.~(\ref{be5}) can be extended to $\infty$. In this limit, 
\beq
\label{smallT}
\frac{1}{\tau_{\text{tr}}} =  \frac{\alpha}{\pi^2} \frac{T^2}{E_0} \bigg( \frac{T}{
T_{\text{BG}}} \bigg)^3 
\eeq
with $E_0$ given in Eq.~(\ref{E0}) and
\bea
\alpha=\int^\infty_0 dx x^3 F(x)\approx 540.0\label{alpha}.
\eea
We see that the scattering rate in this limit scales as $T^5$. A quick inspection of the integral in Eq.~(\ref{alpha}) shows that $x \sim y \sim 1$, i.e.,  $q \sim q_1 \sim T/s$. Note that although the $T^5$ scaling of the scattering rate is the same as for electrons interacting with acoustic phonons via the (single-phonon) deformation-potential mechanism, the  coincidence is accidental.

In the opposite limit of $T \gg T_{\text{BG}}$,  
we assume that typical $x \sim y \ll 1$, which allows us to approximate the Bose function $n_B(x)$ by $1/x$. In this limit, 
\beq
\label{largeT}
\begin{split}
\frac{1}{\tau_{2\text{TO}}} &= T^2 \bigg( \frac{T}{T_\text{BG}} \bigg)^3 \frac{2 m^* g_2^2
\Omega_0^4}{s^4 (2\pi)^5} \int_0^{T_{\text{BG}}/T} dx x^3 \int_0^\infty dy y \int_{-1}^1 d(\cos\theta_{\bx\by}) \bigg( 1 + \frac{(\bx \cdot \by - y^2)^2}{y^2 |\bx - \by|^2} \bigg) \frac{1}{y} \frac{1}{|\bx - \by|^2} \\
&= \frac{T^2}{E_0},
\end{split}
\eeq
which reproduces the first term in Eq.~(6) of the MT. We see that typical $x \sim y \sim T_\text{BG}/T \ll 1$, and thus our assumption of $x \sim y \ll 1$ is justified. 

\begin{figure}{H}
\centering
\includegraphics[scale=0.5]{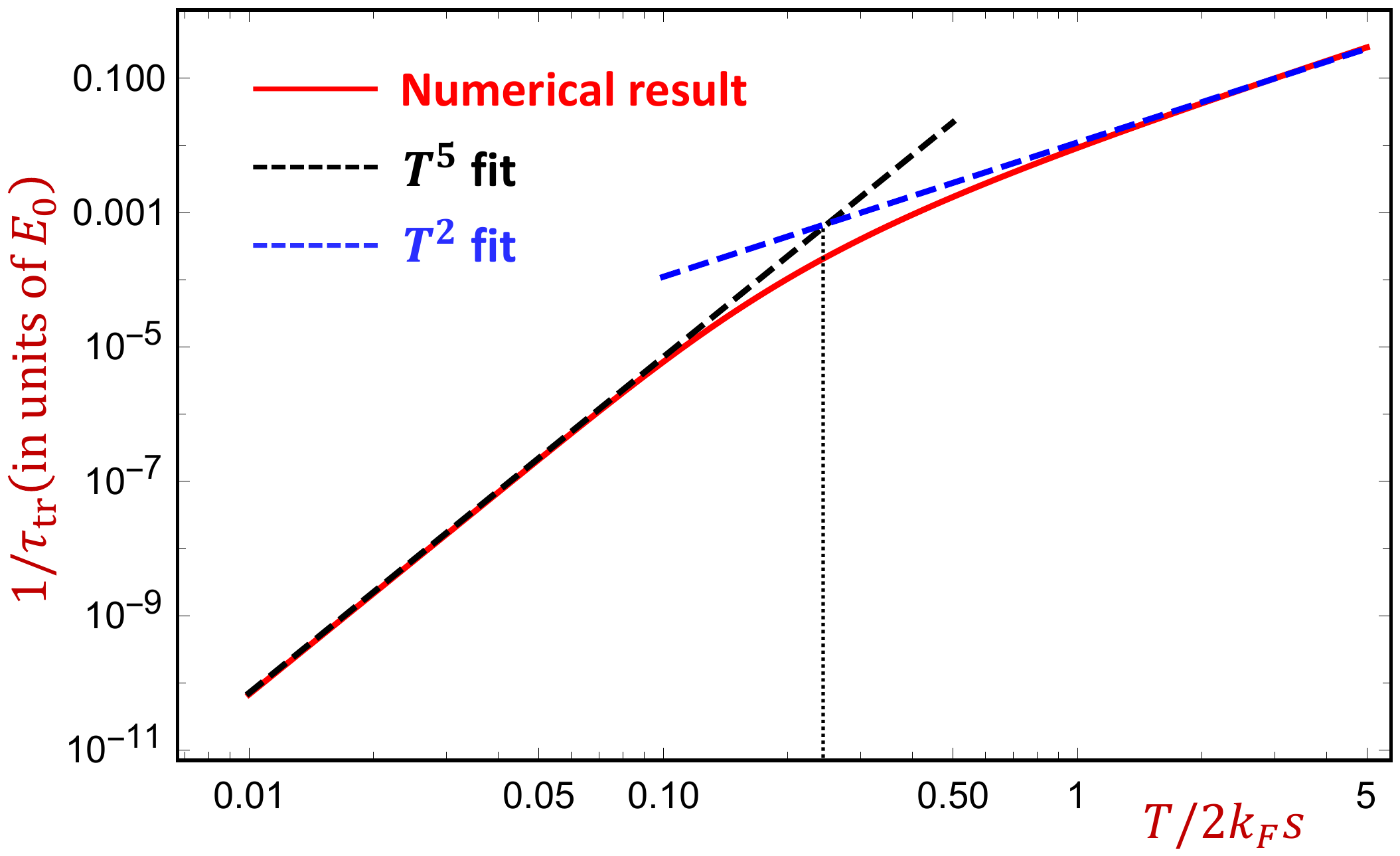}
\caption{\label{crossover} Crossover between the $T^2$ and $T^5$ scaling regimes of the transport scattering time (in units of $E_0$). The $T^2$ and $T^5$ asymptotes intersect at $T_\text{cr} \approx 0.25T_\text{BG}$}
\end{figure}

To estimate the the crossover temperature between the $T^5$ and $T^2$ regime, we equate Eqs.~\eqref{smallT} and \eqref{largeT} and obtain the crossover temperature as $T_{\text{cr}} \approx (\pi^2/\alpha)^{1/3}\approx 0.26 T_{\text{BG}}$. Calculating the integral in Eq.~(\ref{be5}) numerically and defining the crossover temperature as the intersection point of the $T^2$ and $T^5$ asymptotes, we obtain a very close value of
$T_{\text{cr}}\approx 0.25T_{\text{BG}}$, see Fig.~\ref{crossover}.

Now, we turn to the regime $T \ll ms^2$ when the $q/2k_F$ term in the arguments of all the delta functions in Eq.~\eqref{delta} can be neglected compared to the two last term. A typical value of $\cos\theta_{\bp\bq}$ in this regime is on the order of $s/v_F$. Therefore, one factor of $q$ is replaced by $s/v_F$ and the temperature dependence of the scattering rate is now $T^4$ instead of $T^5$. After integration over $\cos\theta_{\bp\bq}$ and switching to a dimensionless form,  Eq.~\eqref{be2} can be written  as
\beq
\label{be6}
\begin{split}
\delta \sigma =& - T^4 \frac{
\beta g_2^2 e^2 N_F k_F \tau_i^2 \Omega_0^4}{
768\pi^5 m^{*2} v_F^2 s^5
},
\end{split}
\eeq
where
\beq
\begin{split}
 \beta=&\int_0^\infty dx x \int_0^\infty dy y \int_{-1}^1 d(\cos\theta_{\bx\by}) \bigg( 1 + \frac{(\bx \cdot \by - y^2)^2}{y^2 |\bx - \by|^2} \bigg) \times \\
&\hspace{0.5cm} \times \bigg[ \frac{(y + |\bx - \by|)^2}{|\bx - \by|} \Big( n_B(y + |\bx - \by|)(1 + n_B(y)) (1 + n_B(|\bx - \by|)) + (1 + n_B(y + |\bx - \by|)) n_B(y) n_B(|\bx - \by|) \Big) \\
&\hspace{0.7cm} + \frac{(y - |\bx - \by|)^2}{|\bx - \by|} \Big( n_B(y - |\bx - \by|)(1 + n_B(y)) n_B(|\bx - \by|) + (1 + n_B(y - |\bx - \by|)) n_B(y) (1 + n_B(|\bx - \by|)) \Big) \bigg]\nn\\
&\approx 67.
\end{split}
\eeq
Finally, the scattering rate for this limit reads
\beq
\frac{1}{\tau} =
\frac{\beta}{\pi^2}  \frac{s}{v_F}  \frac{T^2}{E_0} \left(\frac{T}{T_{\text{BG}}}\right)^2.
\eeq

The crossover temperature between $T^4$ and $T^5$ scaling is $T^*= (\beta/\alpha)(s/v_F)T_{\text{BG}}
\approx 
0.12 (s/v_F)
 T_{\text{BG}}\ll  T_{\text{BG}}$. 

Our final result for temperature dependence of resistivity controlled by two-phonon interaction with massless TO mode is as follows
\beq
\label{cross}
\begin{split}
\rho \propto \begin{cases} T^4, & \,\, T < 
0.12(s/v_F)T_{\text{BG}}  \\
T^5, & \,\, 
0.12(s/v_F) T_{\text{BG}} < T < 0.25
T_{\text{BG}} \\
T^2, & \,\, T > 0.25
T_{\text{BG}}.
\end{cases}
\end{split}
\eeq
For $s=6.6\times 10^5$\,cm/s and $n=4.0\times 10^{17}\cm$, $0.12(s/v_F)T_{\text{BG}}\approx 1.3$\,K and  $0.25 T_{\text{BG}}\approx 6$\,K.

Now we recall that the gap in the spectrum for the lowest TO mode (the $E_u$ mode \cite{Vogt1995,Hehlen1999,yamanaka:2000}) is $\omega_0\approx 1$\ meV $\approx 11$\,K in STO, which is higher than the crossover between the $T^2$ and $T^5$ regimes, estimated above to be  6 K. This means that, as $T$ is lowered, the $T^2$ regime is followed by an exponential freezout ($\rho\propto e^{-\omega_0/T}$) rather than the $T^5$ regime. This is the reason for ignoring the $T^5$ regime in the MT.

Two remarks are now in order. First, the analysis above ignores the interference effects between 
electron-phonon and electron-impurity scattering,\cite{Koshino1960a, *Koshino1960b, *Taylor1962,reizer:1987} which become important when the phonon wavelength becomes longer that the impurity mean free path, i.e., for  $T\ll T_{\text{int}}\ll s/v_F\tau_i$. Our results can be extended into the low-temperature range provides that the lowest crossover temperature in  \eqref{cross}, i.e., $ms^2$, is still higher than $T_{\text{int}}$. The corresponding condition is $ T_{\text{BG}} \gg 1/\tau_i$. Second, the $T^4$ regime has no analog for single acoustic-phonon scattering because the $s/v_F$ term in this case cancels out exactly between the emission and absorption contributions to the collision integral.

\section{Comparison with experiment}
In this section, we give a short description of our fitting procedure to the available experimental data.  We are provided \cite{kamran_thanks} with the data for the dielectric constant ($\ve_0$) of undoped STO at temperatures from 0.3 K to 300 K and for the resistivity of oxygen-reduced STO with carrier density $n = 4 \times 10^{17} \text{cm}^{-3}$ at temperatures from 2 K to 300 K. The dielectric constant data is used to parametrize the temperature dependent gap, $\omega_0(T)$, of the TO phonon dispersion, $\omega^2(q,T) = \omega_0^2(T) + s^2 q^2$ via an empirical   Lyddan-Sachs-Teller (LST) relation, $\ve_0(q,T)=\Omega_0^2/\omega^2_0(q,T)$ with $\Omega_0=194.4$\,meV (Ref.~\onlinecite{Yamada}). 

\begin{figure}
\centering
\includegraphics[scale=0.4]{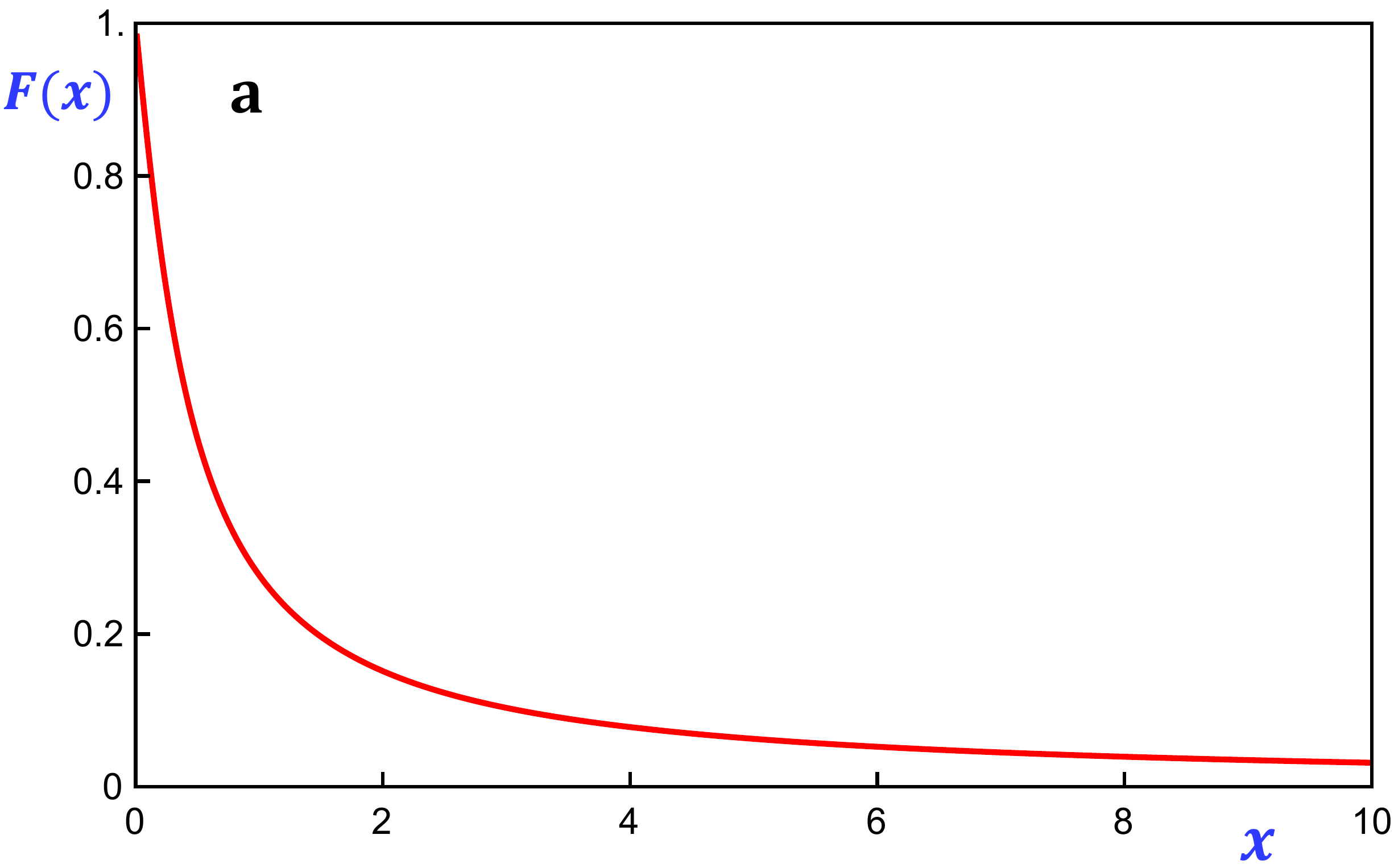}
\includegraphics[scale=0.4]{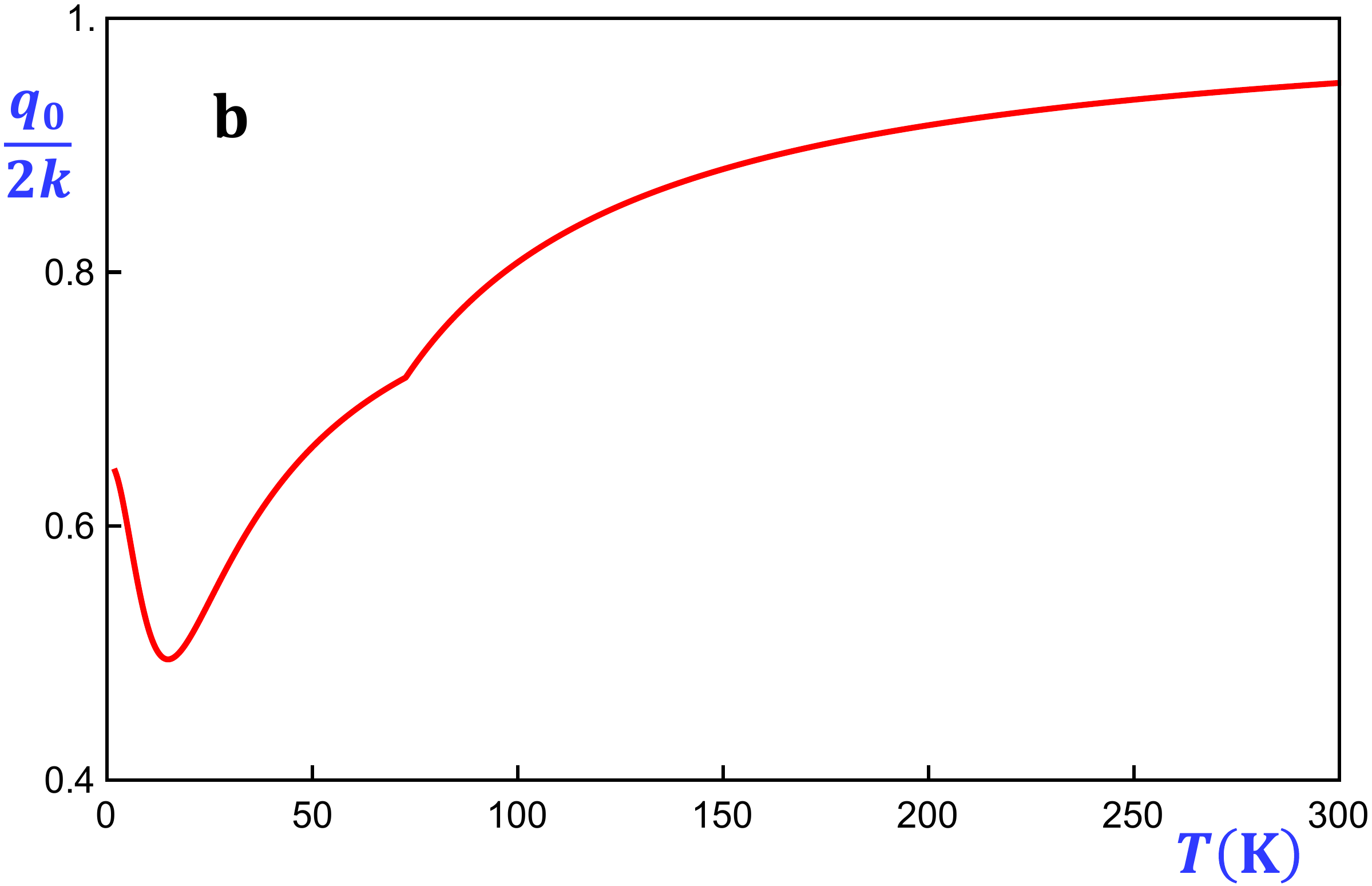}
\includegraphics[scale=0.4]{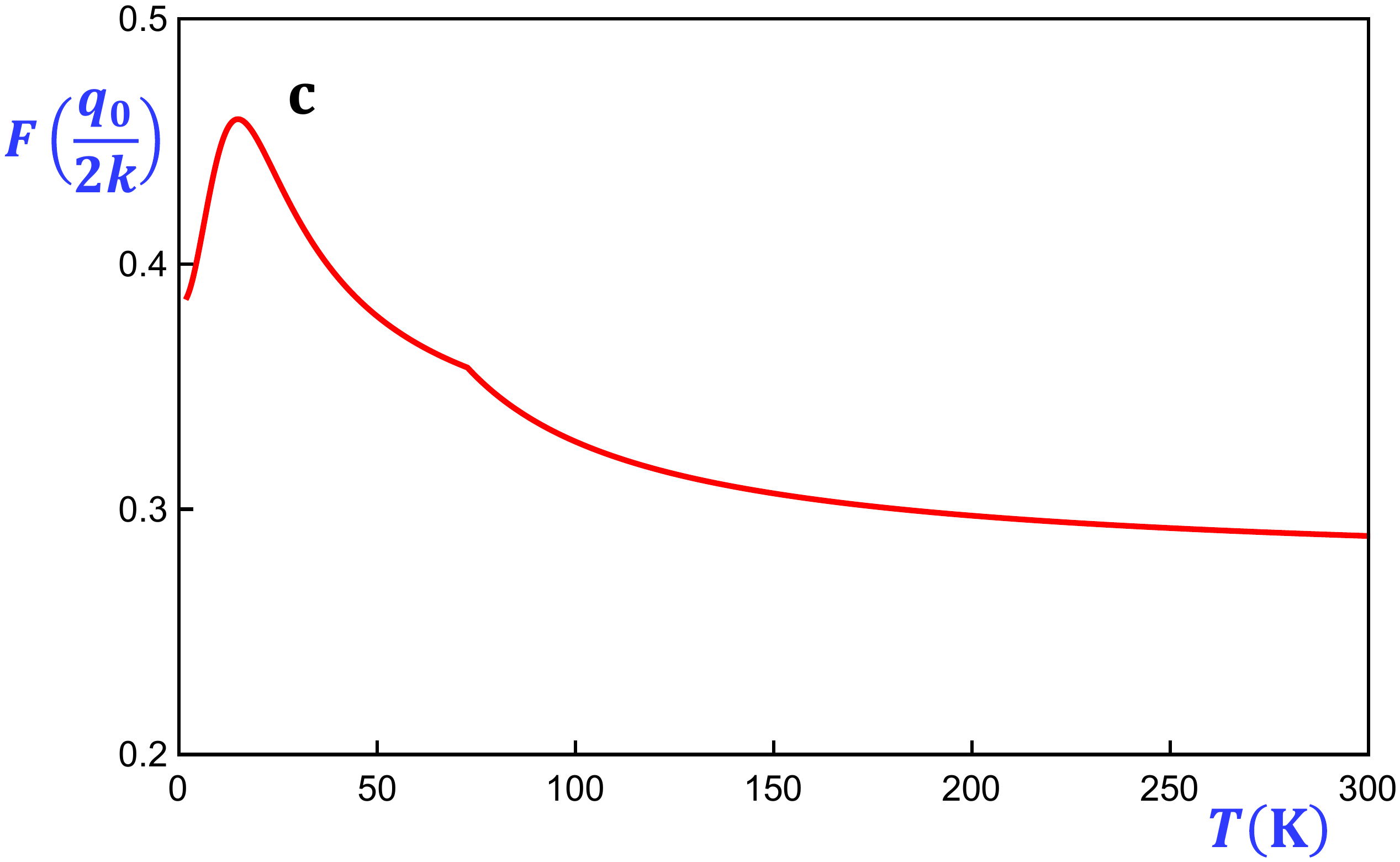}
\caption{\label{dimless_parm}a) Scaling function $F(x)$ in Eq.~(\ref{tr_mass}). b) Dimensionless argument, $q_0/2k(T)$,   of the scaling function in Eq.~(\ref{mass}) as a function of temperature. The $T$-dependence of $q_0$ is calculated using the data for $\ve_0(T)$. A kink at $T\approx 70$\,K is an artifact resulting from combining two sets of data and is to be ignored. c) Variation of the scaling function in Eq.~(\ref{tr_mass}) with $T$.}
\end{figure}

For massive TO phonons, Eq.~(\ref{tr_rate}) for the the transport scattering rate in the quasiparticle regime is replaced by
\beq
\label{tr_mass}
\frac{1}{\tau(k)} = \frac{T^2}{E_0} F\left( \frac{q_0}{2k} \right),
\eeq
where $q_0(T) = \omega_0(T)/s$ is the characteristic phonon momentum and
\beq
\label{mass}
F(x) = \frac{4}{\pi^2} \int_0^1 dz z^3 \int_0^\infty du u^2 \int_{-1}^1 dp \frac{1}{u^2 + x^2} \frac{1}{(\bu - z\hat{z})^2 + x^2} \left[ 1 + \left( \frac{\bu \cdot (z \hat{z} - \bu)}{u (z \hat{z} - \bu)} \right)^2 \right]
\eeq
with $z = q/2k$, $u = q_1/2k$, $p = \cos(\angle\{ \bu, \hat{z} \})$ and $x = q_0/2k$. Function $F(x)$ is plotted in Fig.~\ref{dimless_parm}a.  At $x=0$ we have $F(0)=1$, which reproduces the result for the massless case. For $x\gg 1$, $F(x)\propto 1/x$. Both the numerator and denominator in the quotient $q_0/2k$ vary with temperature
but in such a way that these two dependences partially cancel each other.\cite{levanyuklong,levanyukshort}
Indeed, when calculating the resistivity, $\tau(k)$ is being averaged over thermal distribution. Consequently,  $k$ can be replaced by an rms electron momentum, $k^2(T)=\int d\xi N(\xi) k^2 n_F'/\int d\xi N(\xi)n_F'$, which varies from $k(T)\approx k_F$ for $T\ll E_F$ to $k(T)=\sqrt{2m^*T}$ for $T\gg E_F$. On the other hand, $q_0$ saturates at a $T$-independent value at $T\to 0$ and varies as $\sqrt{T}$ at high $T$, when $\ve_0(T)$ follows a Curie dependence, $\ve_0(T)\propto 1/(T-T_c)\approx 1/T$. We see that the argument of function $F$ is independent of $T$ both at low and high $T$. For intermediate temperatures, the cancellation is not complete but, as Fig.~\ref{dimless_parm}b shows, the variation of $q_0/2k$ with $T$ over a wide temperature range is not significant. A concurrent variation of $F$ with $T$ is shown in Fig.~\ref{dimless_parm}c.

In principle, the resistivity for massive phonons in the non-quasiparticle regime can be obtained from solving a self-consistent Dyson equation, similar to the massless case considered in Sec.~\ref{sec:nqp} However, the non-quasiparticle regime sets in at higher $T$, where the effect of finite gap in the phonon spectrum is less pronounced. 
Therefore, we ignore this complication and use the massless result $\rho\propto T^{3/2}$ to describe the high-$T$ part. Over the whole range of tempartures, the resistivity from 2TO mechanism is described by an interpolating formula
\beq
\label{interp}
\rho_{2TO}(T)= \frac{m^*}{ne^2}\frac{T^2}{E_0} \frac{1}{1 + a \sqrt{\frac{T}{E_0}}} F\left( \frac{q_0}{2k(T)} \right),
\eeq
where  $a = 4\sqrt{2}/C\pi^{3/2}=0.18$ with $C$ defined in Eq.~(\ref{C}). This formula
reproduces the quasiparticle result for massive phonons for $T\ll E_0$ and scales as $T^{3/2}$ for $T\gg E_0$. Since the scaling function in Eq.~(\ref{tr_mass}) is essentially independent of $T$ for $T>150$\,K, it does not affect the dependence of $\rho$ on $T$ at higher $T$. 
To account for faster than $T^2$ variation of the observed resistivity with $T$ at $T\gtrsim 100$\,K,
we include Fr{\"o}hlich-like scattering from the  58 meV LO mode in STO, treating the  Fr{\"o}hlich coupling constant, $\alpha$, as a fitting parameter. Another fitting parameter is the 2TO coupling constant, $g_2$, which enters through $E_0$ (cf. Eq.~(\ref{E0})).
Then the final fitting formula for the resistivity  reads  
\beq
\rho - \rho_0 = 
\rho_{2TO}(T)+ A(\alpha) e^{-\frac{\omega_{LO}}{T}},
\eeq
where $\rho_0$ is the residual resistivity,  $A(\alpha)$ is the Low-Pines function \cite{lowpines}
\beq
A(\alpha) = \frac{2\omega_\text{LO} m^*}{n e^2} \alpha \left( 1 + \frac{\alpha}{6} \right)^3,
\eeq
and $\omega_{LO}=58$\,meV. The resultant fit is shown in Fig.~2 of the MT. The best fit is achieved for $g_2 = 0.92 a_0^3$, where $a_0=3.9\AA$ is the STO lattice constant, and $\alpha = 2.38$, respectively.

\bibliography{STOref_4}